\documentclass[a4paper,10pt,epsf]{article}

\usepackage{amssymb,amsmath,amsthm}
\usepackage{textcomp}
\usepackage{latexsym}
\usepackage[english]{babel} 
\usepackage[latin1]{inputenc}  
\usepackage[T1]{fontenc}   
\usepackage[all]{xy}
\usepackage{multicol}
\usepackage{amsfonts}
\usepackage{graphics}
\usepackage{epsfig}
\usepackage{graphicx}
\usepackage[all]{xy}

\newcommand{\color}[6]{}

\def\CC{\mathbb{C}}
\def\R{\mathbb{R}}
\def\C{\mathbb{C}}
\def\ZZ{\mathbb{Z}}
\def\NN{\mathbb{N}}

\def\PP{\mathbb{P}}

\newcommand{\incl}{\ar@{^{}-}}

\def\gG{\mathfrak{g}}

\def\S{{\mathfrak  S}}

\newtheorem{prop}{Proposition}[section]

\newtheorem{lemma}{Lemma       }
\newtheorem{def }{Definition  }[section]
\newtheorem{theorem}{Theorem}
\newtheorem{ex}{Example     }[section]
\newtheorem{rem}{Remark    }[section]

\begin{document}

\title{Geometric descriptions of entangled states by auxiliaries varieties}
\author{Fr\'ed\'eric Holweck\footnote{frederic.holweck@utbm.fr, 
Laboratoire M3M,
Universit\'e de Technologie de Belfort-Montb\'eliard.}, 
Jean-Gabriel Luque\footnote{jean-gabriel.luque@univ-rouen.fr, Universit\'e de Rouen, Laboratoire d'Informatique, du Traitement de l'Information et des Syst\`emes (LITIS), Avenue de l'Universit\'e - BP 8
6801 Saint-\'etienne-du-Rouvray Cedex, FR } and Jean-Yves Thibon\footnote{jean-yves.thibon@univ-mlv.fr, \ Institut Gaspard Monge
Universit\'e de Marne-la-vall\'ee,
77454 Marne-la-Vall\'ee Cedex 2}}

\maketitle

\begin{abstract}
 The aim of the paper is to propose  geometric descriptions of multipartite entangled states using algebraic geometry. In the context of this paper, 
geometric means each stratum of the Hilbert space, corresponding to an entangled 
state, is an open subset of an algebraic variety built by classical geometric constructions (tangent lines, secant lines) from the set of separable states. 
In this setting we describe well-known classifications  of multipartite entanglement such as $2\times 2\times(n+1)$, for $n\geq 1$,
 quantum systems and a new description with the $2\times 3\times 3$ quantum system.
 Our results complete the approach of A. Miyake and makes stronger connections with recent work of algebraic geometers. 
Moreover for the quantum systems detailed in this paper we propose an algorithm, 
based on the classical theory of invariants, 
to decide to which subvariety of the Hilbert space a given state belongs.
\end{abstract}

\section{Introduction}
Let $\mathcal{H}=\CC^{n_1}\otimes\CC^{n_2}\otimes\dots\otimes\CC^{n_k}$ be the Hilbert space of states of $k$ particles. 
Denote by $|j_i\rangle$
 a basis of $\CC^{n_i}$ with $0\leq j_i\leq n_i-1$. Any state $|\Psi\rangle\in \mathcal{H}$ can be written as 
\[|\Psi\rangle=\sum_{1\leq i\leq k}\sum_{0\leq j_{i}\leq n_{i}-1} A_{j_{1}j_{2}\dots j_{k}}|j_1\rangle\otimes\dots\otimes|j_k\rangle\]
 where $|j_1\rangle\otimes\dots\otimes|j_k\rangle$ is the standard basis of $\mathcal{H}$. That basis will be denoted latter on by $|j_1\dots j_k\rangle$.
The Hilbert space $\mathcal{H}$ is an irreducibe representation (for its natural action \cite{F-H}) of 
the semi-simple Lie group $G=SL(n_1,\CC)\times \dots\times SL(n_k,\CC)$. 
In the framework of Quantum Information Theory (QIT), $G$ is the group of reversible stochastic local quantum 
operations assisted
by classical communication (SLOCC, see \cite{My3}), and two states will be considered as SLOCC equivalent if they are interconvertible by the action of $G$,
\[|\Psi\rangle\sim_{\text{SLOCC}} |\Phi\rangle \Leftrightarrow |\Psi\rangle=g|\Phi\rangle,|\Phi\rangle=g^{-1}|\Psi\rangle, \text{ with }g=(g_1,\dots,g_k)\in SL(n_1,\CC)\times \dots\times SL(n_k,\CC)\]


\noindent Nonzero scalar multiplication has no incidence on a state $|\Psi\rangle$, therefore we can consider states as points in
 the projective space $\PP(\mathcal {H})$ and SLOCC equivalent states will
correspond to points in the same $G$-orbit.
The representation
$\mathcal{H}$ of $G$ has a unique highest weight vector which can be chosen to be $v=|0\dots0\rangle$ (it corresponds to a choice of orientation for the weight lattice see \cite{F-H}). 
The orbit $G.v\subset \mathcal{H}$ is the unique closed orbit for the action of $G$ on $\mathcal{H}$ and it
 defines a smooth algebraic variety
after projectivization $X=\PP(G.v)\subset \PP(\mathcal{H})$. 
This variety $X$ is known as the Segre embedding of the product
 of the projective spaces $\PP^{n_i-1}$, and is the image of the  map (see \cite{Ha}):
\[\begin{array}{cccc}
   \phi: & \PP(\CC^{n_1})\times\PP(\CC^{n_2})\times\dots\times\PP(\CC^{n_k}) & \to & \PP(\CC^{n_1}\otimes\CC^{n_2}\otimes\dots\otimes\CC^{n_k})\\
             & ([v_1],[v_2],\dots,[v_k]) & \mapsto & [v_1\otimes v_2 \otimes \dots \otimes v_k] 
  \end{array}\]
where $v_i$ is a vector of $\CC^{n_i}$ and $[v_i]$ the corresponding point in $\PP^{n_i-1}=\PP(\CC^{n_i})$.
The variety $X=\PP(G.v)=\phi(\PP(\CC^{n_1})\times\PP(\CC^{n_2})\times\dots\times\PP(\CC^{n_k}))$ will be denoted by
\[X=\PP^{n_1-1}\times\dots\times\PP^{n_k-1}\subset \PP(\mathcal{H})\] 

\noindent From the QIT point of view \cite{brody1,hey}, the variety $X$ is the set of separable states in $\PP(\mathcal{H})$.\\

\noindent In this paper we cover some examples ($2\times 2\times(n+1)$, with $n\geq 1$, and $2\times3\times3$ quantum systems) 
of  classifications of multipartite entanglement.
 We describe the entangled states by auxiliary varieties, i.e. varieties obtained
 from $X$ by geometric constructions and we propose an algorithm to distinguish between the different states.
The constructions of auxiliary varieties by secant and tangent lines are explained in Section \ref{aux} and some classical results 
of algebraic geometry on the dimension of those varieties are recalled.
In Section \ref{strat} we give the geometric descriptions of the orbit closures for the 
quantum systems $2\times 2\times(n+1)$, $n\geq 1$, (Theorem \ref{identification}) and $2\times 3\times 3$ (Theorem \ref{identification2}).
The description of the (projectivized) Hilbert space 
by different classes of entanglement corresponds to a stratification 
of the ambient space by algebraic varieties with natural geometric inclusions among those 
varieties (the ``onion like'' structure of \cite{My}). 
Those inclusions are detailed in Figures \ref{222onion}, \ref{223onion}, \ref{22nonion} and \ref{233onion}.
The proofs of  Theorem \ref{identification} and Theorem \ref{identification2} 
are based on dimension counts to 
identify the algebraic (auxiliary) varieties with the corresponding orbit closures. Those orbits are known from their representatives \cite{Pav}. 
A geometric description of $2\times 2\times(n+1)$, with $n\geq 1$, quantum systems was already established in \cite{My2} using the concepts of projective duality, hyperdeterminants 
and singular locus of hyperdeterminants. In Section \ref{dual} we
recover the results of \cite{My2} directly from Theorem \ref{identification}. 
We prove that Theorem \ref{identification} and the results of \cite{My2} are dual 
to each other and thanks to Theorem \ref{identification}
we obtain more details in the geometric description of the singular locus of the dual variety.
In Section \ref{algo} we introduce classical invariant theory in the context of hypermatrices. We obtain
 an algorithm to decide to which strata (variety) of the 
ambient space belongs a given state $|\Psi\rangle$.
In other words our method allows us to identify the orbit of a given state $|\Psi\rangle$. 
This section is mainly based on classical invariant theory techniques but we relate  part of 
the information obtained by those techniques with  geometric descriptions.
Finally  we mention recent works of algebraic geometers which we
 believe should help to provide deeper understanding of multipartite entanglement.

\subsection*{Notations}

\noindent We work throughout with algebraic varieties over the field
$\CC$ of complex numbers. In particular we denote by $V$ a complex
vector space of dimension $N+1$ and $X^n\subset \PP(V)=\PP^{N}$ is a
complex projective nondegenerate variety (i.e. not contained in a hyperplane)
of dimension $n$. 
Given $x$ a smooth point of $X$, we denote by $T_x
X$ the intrinsic tangent space, $\tilde{T}_x X$ the embedded tangent
space, of $X$ at $x$ (see \cite{Lan}). The notation $\hat{X}\subset V$ (resp. $\widehat{T}_x X$) will denote the cone over 
$X$ (resp. over $\tilde{T}_x X$) and $[v]\in \PP(V)$ will denote the projectivization of a vector $v\in V$. 
The dimension of the variety, $\text{dim}(X)$, is the dimension of the tangent space at a smooth point.
We say $x\in X$ is a {\em
  general point} of $X$ in the sense of the Zariski topology. The locus of
smooth points of $X$ is denoted by $X_{smooth}$ and the locus of
singular points by $X_{sing}$. 

\section{Join varieties}\label{aux}

Let $x$ and $y$ be two points of $\PP^N$, the secant line $\PP^1_{xy}$ is the unique line in $\PP^N$ containing $x$ and $y$. We define the join of two varieties
$X$ and $Y$ to be the (Zariski) closure of the union of the secant lines with $x\in X$ and $y\in Y$:
\[J(X,Y)=\overline{\bigcup_{x\in X, y\in Y, x\neq y} \PP^{1}_{xy}}\]

\noindent Suppose $Y\subset X$ and let $T^{\star}_{X,Y,y_0}$ denote the union of $\PP^1 _*$'s where $\PP^1 _*$ is 
the limit of $\PP^1_{xy}$ with $x\in X$, $y\in Y$ and $x,y\to y_0\in Y$. 
The union of the $T^{\star}_{X,Y,y_0}$ is defined as the variety of  relative tangent stars of $X$ with respect to $Y$ (see \cite{Z2}):
\[T(Y,X)=\bigcup_{y \in Y}T^{\star}_{X,Y,y} \]
 
\noindent The following result due to F. Zak (\cite{Z2} Chapter I Theorem 1.4) will be useful to analyse the stratification of the ambient space by auxiliary varieties:

\begin{theorem}\label{Zak}
 Any arbitrary irreducible subvariety $Y^n\subset X^m, n\geq 0$ satisfies one of the following two conditions:
\begin{enumerate}
 \item[a)] $\text{dim}(J(X,Y))=n+m+1$ and $\text{dim}(T(X,Y))=n+m$;
 \item[b)] $J(X,Y)=T(X,Y)$.
\end{enumerate}
\end{theorem}

\begin{rem}\rm
 The expected dimension of $J(X,Y)$ is $n+m+1$: there are $n$ degree of freedom for the choice of $y\in Y$, 
$m$ degree of freedom for the choice of $x\in X$ and $1$
degree of freedom for the choice of the point on $\PP^1_{xy}$. Therefore the previous theorem says that if $J(X,Y)$ has the expected dimension then $T(X,Y)$
has also the expected dimension and is distinct from $J(X,Y)$. We will see that dimension calculations will be used later to prove the existence of certain statras.
\end{rem}

\noindent The dimension of $J(X,Y)$ will be calculated with the following famous lemma (see \cite{Lan} Chapter III).
\begin{lemma}{\bf [Teracini's Lemma]}
If $z\in J(X,Y)_{\text{smooth}}$ with $z=x+y$ such that $x\in X_{\text{smooth}}$, $y\in Y_{\text{smooth}}$, then 
\[ \widehat{T}_zJ(X,Y)=\widehat{T}_x X+\widehat{T}_yY\]
 
\end{lemma}

Intersting particular cases arise when $Y=X$.
The s-secant variety of a projective variety $X\subset \PP^{N}$ is the variety $\sigma_s(X)$ defined to be 
the closure of the union of the linear span of $s$-tuples points of $X$
\[\sigma_s(X)=\overline{\bigcup_{x_1,\dots,x_s\in X} \PP^{s-1}_{x_1\dots x_s}}\]
where $\PP^{s-1}_{x_1\dots x_s}$ is a projective space of dimension $s-1$ passing through $x_1,\dots, x_s$.
In other words $\sigma_s(X)=J(X,\sigma_{s-1}(X))$ with $\sigma_1(X)=X$. The variety $\sigma_2(X)$ is often called the secant variety.
For the variety of relative tangent stars we obtain for $Y=X$ the usual tangential variety: \[\tau(X)=T(X,X)=\bigcup_{x\in X_{smooth}} \tilde{T}_x X\]




\begin{ex}\label{bipartite}{\rm
Let $\mathcal{H}=\CC^{m+1}\otimes\CC^{n+1}$, 
and $X=\PP^m\times\PP^n\subset\PP(\mathcal{H})$ be the Segre product of two projective 
spaces, i.e. the unique closed orbit for the action of $SL_{m+1}\times SL_{n+1}$ on $\PP(\mathcal{H})$.
The variety $X$
corresponds to the projectivization of rank one matrices in the projectivization of the space of matrices of size $(m+1)\times (n+1)$.
Therefore the $s$-secant variety $\sigma_s(X)$ is the projectivization of the set of rank less than $s$ matrices (sum of $s$ matrices of rank one).
As noticed in \cite{hey} the stratification of $\PP(\mathcal{H})$ by secant varieties of $\PP^{m}\times \PP^n$ is the stratification by local rank for bipartite quantum systems :
\[\PP^m\times\PP^n\subset \sigma_2(\PP^m\times\PP^n)\subset\dots\subset\sigma_{min(m,n)-1}(\PP^m\times\PP^n)\subset \PP(\mathcal{H})\]
}
\end{ex}

\begin{rem}\rm
 As noticed in \cite{brody1}, a projective line $\PP^1_{xy}$ in the  Hilbert space $\PP(\mathcal{H})$ represents 
all possible superpositions of the states $\hat{x}$, $\hat{y}\in \mathcal{H}$.
\end{rem}

\begin{def }
 Let $X\subset \PP(V)$ be an irreducible variety of dimension $n$. The $s$-secant variety $\sigma_s(X)$ is said to be nondefective if either $\text{dim}(\sigma_s(X))=sn+s-1$ or $\sigma_s(X)=\PP(V)$.
\end{def }

\noindent A direct consequence of Theorem \ref{Zak} is the following proposition :

\begin{prop}\label{secant-tangent}
 Let $X^n\subset \PP(V)$ be a nondegenerate variety. Let us assume that the $k$-th secant variety is nondefective and does not fill the ambient space. 
Then we have the following filtration with the given dimensions:
\[\underbrace{X}_{\text{dim}=n}\subset \underbrace{\tau(X)}_{=2n}\subset \underbrace{\sigma_2(X)}_{=2n+1}\subset \underbrace{T(X,\sigma_2(X))}_{=3n+1}\subset \underbrace{\sigma_3(X)}_{=3n+2} \subset \underbrace{T(X,\sigma_3(X))}_{=4n+2}\subset \dots\subset \underbrace{\sigma_{k}(X)}_{=kn+k-1} \subset\PP(V)  \]

\end{prop}

\proof If $\sigma_k(X)$ is nondefective, i.e. is of dimension $kn+k-1$  then one knows from Theorem \ref{Zak} that $T(X,\sigma_{k-1}(X))$ is of dimension $kn+k-2$. 
Moreover Theorem \ref{Zak} ensures us in this case  $\text{dim}(\sigma_k(X))= \text{dim}(\sigma_{k-1}(X))+\text{dim}(X)+1$, thus $\text{dim}(\sigma_{k-1}(X))=(k-1)n+(k-1)-1$, i.e. $\sigma_{k-1}(X)$ is nondefective and we apply the same argument inductively $\Box$.

\begin{rem}\rm
 Proposition \ref{secant-tangent} gives {\em a priori} a filtration by secant and tangential varieties and explain part of the ``onion like'' structure described in \cite{My}. 
As $X$ is a Segre product of projective spaces, Definition \ref{jpair} shows that there are others intermediate 
auxiliary varieties which  will appear in the filtration of the 
ambient space.
\end{rem}

\noindent We will need the following definition to identify subvarieties of $\sigma_2(X)$ when $X$ is a Segre product of irreducible varieties.
 In Section $\ref{dual}$ Definition \ref{jpair} will also be used to describe the singular locus of the dual variety.
%
%

\begin{def }\label{jpair}
 Let $Y_i\subset\PP^{n_i}$, with $1\leq i\leq m$ be  $m$ nondegenerate varieties and let us consider 
$X=Y_1\times Y_2\times\dots\times Y_m\subset \PP^{(n_1+1)(n_2+1)\dots(n_m+1)-1}$ the corresponding Segre product. 
For $J=\{j_1,\dots,j_k\}\subset \{1,\dots,m\}$, a $J$-pair of points of $X$ will be a pair $(x,y)\in X\times X$ such that 
$x=[v_1\otimes v_2\otimes\dots\otimes{v_{j_1}}\otimes v_{j_1+1}\otimes\dots\otimes{v_{j_2}}\otimes\dots\otimes{v_{j_k}}\otimes\dots\otimes v_{m}]$
  and 
$y=[w_1\otimes w_2\otimes\dots\otimes {v_{j_1}}\otimes w_{j_1+1}\otimes\dots\otimes {v_{j_2}}\otimes\dots\otimes{v_{j_k}}\otimes\dots\otimes w_{m}]$, 
i.e. the tensors $\hat{x}$ and $\hat{y}$ have the same components for the indices in $J$.
 
The $J$-subsecant variety of $\sigma_2(X)$ denoted by $\sigma_2(Y_1\times\dots \times\underline{Y}_{j_1}\times \dots\times \underline{Y}_{j_k}\times \dots\times Y_m)\times Y_{j_1}\times Y_{j_2}\times\dots\times Y_{j_k}$
 is the closure of the union of line
 $\PP^1_{xy}$ with $(x,y)$ a $J$-pair of  point:

\[\sigma_2(Y_1\times\dots \times\underline{Y}_{j_1}\times \dots\times \underline{Y}_{j_k}\times \dots\times Y_m)\times Y_{j_1}\times Y_{j_2}\times\dots\times Y_{j_k}
=\overline{\cup_{(x,y)\in X\times X, (x,y) J-\text{pair of points}} \PP_{xy} ^1}\] 
\end{def }

\begin{rem}\rm
 The underlined varieties in the notation of the $J$-subsecant varieties correspond to the common components for the points which define 
a  
$J$-pair. Roughly speaking those components are the ``common factor'' of $x$ and $y$ in
the decomposition of 
$z=x+y\in  \sigma_2(Y_1\times\dots \times\underline{Y}_{j_1}\times \dots\times \underline{Y}_{j_k}\times \dots\times Y_m)\times Y_{j_1}\times Y_{j_2}\times\dots\times Y_{j_k}$. 
For instance when we consider the $\{1\}$-subsecant (respectively the $\{m\}$-subsecant) 
variety we can indeed factorize the first (respectively the last) component and we have the equality $\sigma_2(\underline{Y}_1\times Y_2\times\dots \times Y_m)\times Y_1=Y_1\times \sigma_2(Y_2\times\dots\times Y_m)$.
\end{rem}

\begin{rem}\rm
 For $J=\emptyset$, the $J$-subsecant variety is $\sigma_2(X)$.
\end{rem}

\section{Stratification of the multipartite entangled states by the tangential, secant and join varieties}\label{strat}
We now state with Theorem \ref{identification} and Theorem \ref{identification2} our geometric descriptions 
of entangled states for quantum systems of type $2\times 2\times(n+1)$, $n\geq 1$,
 and $2\times 3\times 3$ by join, secant and tangential varieties.
The orbit closures will be denoted $\overline{\mathcal{O}}_{*}$ where the subscript is either a representative or a roman number to identify the orbit.

\begin{figure}[!h]
\[\xymatrix{&  \PP^7&  \\ 
&\tau(\PP^1\times \PP^1 \times \PP^1)\incl[dr]\incl[u] &  \\
    \sigma_2(\PP^1\times \PP^1)\times \PP^1       \incl[ru] \incl[rd] & \PP^1\times\sigma_2(\PP^1\times \PP^1)\incl[u] \incl[d]& \sigma_2(\PP^1\times\underline{\PP^1}\times \PP^1)\times \PP^1 \\
 & \PP^1\times \PP^1 \times \PP^1 \incl[ur]&  \\
}\]

\caption{Stratification of the ambient space for the $2\times 2\times 2$ quantum system}\label{222onion}
\end{figure}
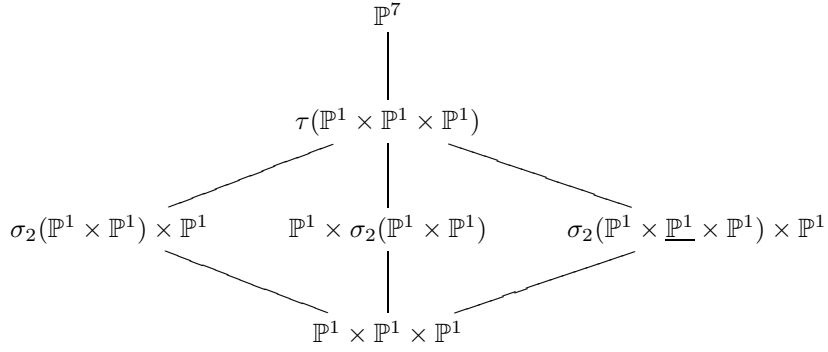

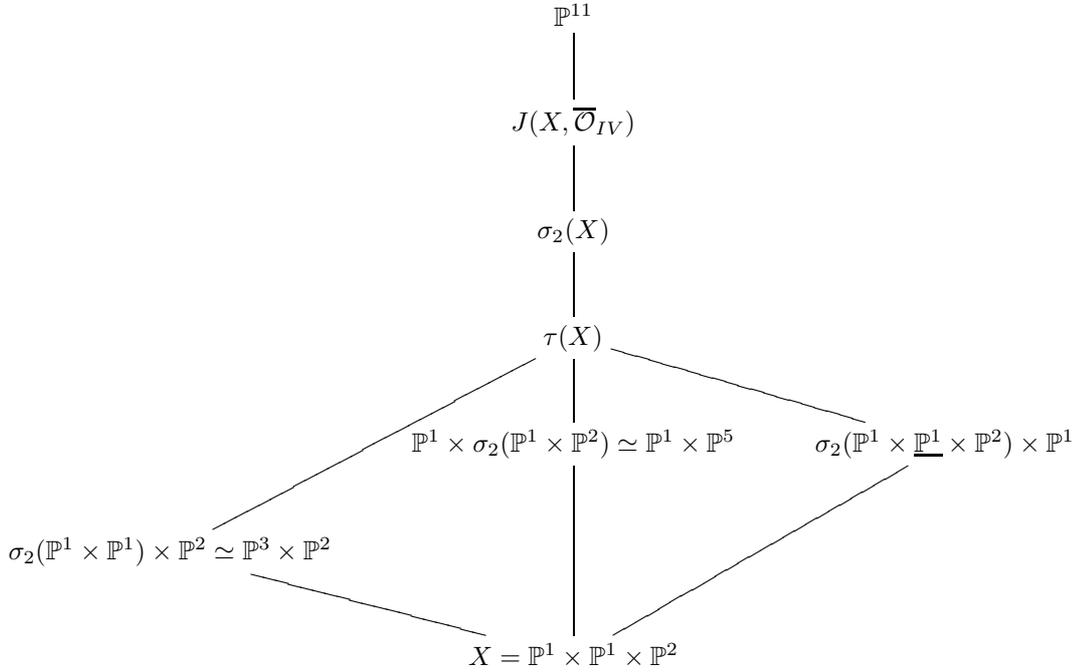
\begin{figure}[!h]
\[\xymatrix{ & \PP^{11} & \\
& J(X,\overline{\mathcal{O}}_{IV}) \incl[u] & \\
& \sigma_2(X) \incl[u] & \\
& \tau(X) \incl[u] & \\
&  \PP^1\times\sigma_2(\PP^1\times \PP^2)\simeq \PP^1\times\PP^{5}\incl[u] & \sigma_2(\PP^1\times \underline{\PP^1}\times \PP^2)\times \PP^1 \incl[ul]\\
\sigma_2(\PP^1\times \PP^1)\times\PP^2\simeq \PP^3\times \PP^2\incl[uur] &  & \\
& X=\PP^1\times \PP^1 \times \PP^2 \incl[uu] \incl[ruu] \incl[ul] & \\
}\]
\caption{Stratification of the ambient space for the $2\times 2\times 3$ quantum system}\label{223onion}
\end{figure}

\begin{theorem}\label{identification}
 For quantum systems of type $2\times 2\times(n+1)$, there are $6$ ($n=1)$, $8$ ($n=2$) and $9$ ($n\geq 3$)  SLOCC entangled classes.
Each entangled state corresponds to an open subspace (smooth points) of an algebraic variety build up from 
$X=\PP^{1}\times\PP^{1}\times\PP^{n}$, the unique closed orbit for the action of $G=SL_2\times SL_2\times SL_{n+1}$ on 
$\PP(\mathcal{H})=\PP(\CC^2\otimes\CC^2\otimes\CC^{n+1})$ by join and tangential varieties. The identifications of those algebraic varieties are given 
in Tables \ref{table222}, \ref{table223} and \ref{table22n} and the partial order among them is represented in Figures \ref{222onion}, \ref{223onion}, \ref{233onion}.
\begin{table}[!h]
\begin{center}
\begin{tabular}{c|c|c|c}
\hline
Orbit closure & Normal form (representative) & Variety & Dimension\\
\hline
$\overline{\mathcal{O}}_{VI}$ & $|000\rangle+|111\rangle $&  $\PP^7$ & $7$  \\ 
$\overline{\mathcal{O}}_V$ & $|100\rangle+|010\rangle+|001\rangle$& $\tau(\PP^1\times \PP^1 \times \PP^1)$& $6$ \\
$\overline{\mathcal{O}}_{IV}$ & $|001\rangle+|111\rangle$  &   $\sigma_2(\PP^1\times \PP^1)\times \PP^1$& $4$\\ 
$\overline{\mathcal{O}}_{III}$& $|100\rangle+|111\rangle$ & $\PP^1\times\sigma_2(\PP^1\times \PP^1)$ & $4$ \\
$\overline{\mathcal{O}}_{II}$ & $|010\rangle+|111\rangle$ & $\sigma_2(\PP^1\times\underline{\PP^1}\times \PP^1)\times \PP^1$ & $4$ \\
$\overline{\mathcal{O}}_I$ & $|000\rangle$ &  $\PP^1\times \PP^1 \times \PP^1$ & $3$
\end{tabular}
\caption{Identification of orbit closures and varieties for the $2\times 2\times 2$ quatum system}\label{table222}
\end{center}
\end{table}
\begin{table}[!h]
\begin{center}
\begin{tabular}{c|c|c|c}
\hline
Orbit closure &Normal form (representative) & Variety & Dimension\\
\hline
 $\overline{\mathcal{O}}_{VIII}$&$|000\rangle+|011\rangle+|101\rangle+|112\rangle$&       $\PP^{11}$ &$11$\\
$\overline{\mathcal{O}}_{VII}$ &$|000\rangle+|011\rangle+|102\rangle$& $J(X,\overline{\mathcal{O}}_{IV})$ &$10$ \\
$\overline{\mathcal{O}}_{VI}$ &$|000\rangle+|111\rangle$& $\sigma_2(X)$ & $9$\\
$\overline{\mathcal{O}}_{V}$& $|000\rangle+|011\rangle+|101\rangle$& $\tau(X)$ &$8$ \\
$\overline{\mathcal{O}}_{IV}$&$|000\rangle+|011\rangle$&  $\PP^1\times\sigma_2(\PP^1\times \PP^2)\simeq\PP^1\times \PP^{5}$ & $6$\\
$\overline{\mathcal{O}}_{III}$&$|000\rangle+|101\rangle$   & $\sigma_2(\PP^1\times \underline{\PP^1}\times \PP^2)\times \PP^1$ &$6$\\
$\overline{\mathcal{O}}_{II}$&$|000\rangle+|110\rangle$&$\sigma_2(\PP^1\times \PP^1)\times \PP^2\simeq \PP^3\times \PP^2$ &$5$\\
$\overline{\mathcal{O}}_{I}$&$|000\rangle$& $X=\PP^1\times \PP^1 \times \PP^2$  &$4$
 \end{tabular}
 \caption{Identification of orbit closures and varieties for the $2\times 2\times 3$ quatum system}\label{table223}
\end{center}
\end{table}
\begin{table}[!h]
\begin{center}
\begin{tabular}{c|c|c|c}
\hline
Orbit closure & Normal form (representative) & Variety & Dimension\\
\hline
$\overline{\mathcal{O}}_{IX}$ & $|000\rangle+|011\rangle+|102\rangle+|113\rangle$& $\PP^{4n+3}$ &$4n+3$ \\
 $\overline{\mathcal{O}}_{VIII}$ &$|000\rangle+|011\rangle+|101\rangle+|112\rangle$&        $\sigma_3(X)$ &$3n+5$\\
$\overline{\mathcal{O}}_{VII}$& $|000\rangle+|011\rangle+|102\rangle$& $J(X,\overline{\mathcal{O}}_{IV})$ &$3n+4$ \\
$\overline{\mathcal{O}}_{VI}$ &$|000\rangle+|111\rangle$& $\sigma_2(X)$ & $2n+5$\\
$\overline{\mathcal{O}}_{V}$ &$|000\rangle+|011\rangle+|101\rangle$& $\tau(X)$ &$2n+4$ \\
$\overline{\mathcal{O}}_{IV}$&$|000\rangle+|011\rangle$&  $\PP^1\times\sigma_2(\PP^1\times \PP^n)\simeq\PP^1\times \PP^{2n+1}$ & $2n+2$\\
$\overline{\mathcal{O}}_{III}$&$|000\rangle+|101\rangle$   & $\sigma_2(\PP^1\times \underline{\PP^1}\times \PP^n)\times \PP^1$ &$2n+2$\\
$\overline{\mathcal{O}}_{II}$&$|000\rangle+|110\rangle$&$\sigma_2(\PP^1\times \PP^1)\times \PP^n\simeq \PP^3\times \PP^n$ &$n+3$\\
$\overline{\mathcal{O}}_{I}$&$|000\rangle$& $X=\PP^1\times \PP^1 \times \PP^n$  &$n+2$
 \end{tabular}
 \caption{Identification of orbit closures and varieties for $2\times 2\times(n+1)$ quatum systems $(n\geq 3)$}\label{table22n}
\end{center}
\end{table}
\end{theorem}

\begin{figure}[!h]
\[\xymatrix{ & \PP^{4n+3} & \\
         &  \sigma_3(X) \incl[u] & \\
& J(X,\overline{\mathcal{O}}_{IV}) \incl[u] & \\
& \sigma_2(X) \incl[u] & \\
& \tau(X) \incl[u] & \\
&  \PP^1\times\sigma_2(\PP^1\times \PP^n)\simeq \PP^1\times\PP^{2n+1}\incl[u] & \sigma_2(\PP^1\times \underline{\PP^1}\times \PP^n)\times \PP^1 \incl[ul]\\
\sigma_2(\PP^1\times \PP^1)\times\PP^n\simeq \PP^3\times \PP^n\incl[uur] &  & \\
& X=\PP^1\times \PP^1 \times \PP^n \incl[uu] \incl[ruu] \incl[ul] & \\
}\]
\caption{Stratification of the ambient space for $2\times2\times(n+1)$, $n\geq 3$, quantum system}\label{22nonion}
\end{figure}

\begin{rem}\rm
 After \cite{Dur} the classification of entangled states of the $2\times 2\times 2$ quantum system received a lot of attention 
because it showed for the first time that three qubits  could
be entangled in different inequivalent states. The existence of the so-called GHZ-state and W-state for a 3-qubit system proved in \cite{Dur} is equivalent 
in our description to the existence of the (nondefective) secant and tangential varieties of $\PP^1\times\PP^1\times\PP^1$ which is a classical result of algebraic geometry.
\end{rem}

\begin{rem}\rm
 In \cite{brody2} different geometric characterizations of the 3-qubit system are given. The authors consider the intersection of the 
variety of pure state $X$ with the three plane of symmetric tensors (symmetric states). 
That intersection reduces to the twisted cubic $\mathcal{C}\subset\PP_{\text{sym}}^3$ and allows them to characterize the GHZ-states and W-state in terms of that curve.
\end{rem}

\begin{rem}\rm
The geometric classification of entangled states for $2\times 2\times(n+1)$, $n\geq 2$, systems was proposed in \cite{My,My2} but with a different geometric perspective (i.e. by dual varieties). 
Not all stratas are 
geometrically described in the paper. 
We will make the connection between our identifications and the classifications of \cite{My,My2}  in Section $\ref{dual}$. 
\end{rem}

\noindent The next theorem provides a geometrical description for $2\times 3\times 3$ quantum system.

\begin{theorem}\label{identification2}
For a quantum system of type $2\times 3\times 3$ there are $17$ different SLOCC entangled classes.
Each entangled state corresponds to an open subspace of an algebraic variety constructed from 
$X=\PP^{1}\times\PP^{2}\times\PP^{2}$, the unique closed orbit for the action of $G=SL_2\times SL_3\times SL_3$ on 
$\PP(\mathcal{H})=\PP(\CC^2\otimes\CC^3\otimes\CC^3)$, by join and tangential varieties. The identifications of those algebraic varieties are given 
in Table \ref{table233} and the partial order among them is represented in Figure \ref{233onion}.
\begin{table}[!h]
\begin{center}
 \begin{tabular}{c|c|c|c}
\hline
Orbit closure &Normal form (representative)& Variety & Dimension\\
\hline
$\overline{\mathcal{O}}_{XVII}$&$|000\rangle+|011\rangle+|100\rangle+|122\rangle$& $\PP^{17}$ &$17$ \\
$\overline{\mathcal{O}}_{XVI}$& $|000\rangle+|011\rangle+|101\rangle+|122\rangle$&        $J(X,\tau(X))$ &$16$\\
$\overline{\mathcal{O}}_{XV}$&$|000\rangle+|011\rangle+|022\rangle+|101\rangle+|112\rangle$& $T(X,\tau(X))$ &$15$ \\
$\overline{\mathcal{O}}_{XIV}$&$|000\rangle+|011\rangle+|122\rangle$& $J(X,\PP^1\times\sigma_2(\PP^2\times\PP^2))$ & $14$\\
$\overline{\mathcal{O}}_{XIII}$&$|000\rangle+|011\rangle+|022\rangle+|101\rangle$& $T(X,\PP^1\times\sigma_2(\PP^2\times\PP^2))$ &$13$ \\
$\overline{\mathcal{O}}_{XII}$&$|000\rangle+|011\rangle+|101\rangle+|112\rangle$&  $\sigma_2(\sigma_2(\PP^1\times\underline{\PP^2}\times\PP^2)\times\PP^2)$ & $13$\\
$\overline{\mathcal{O}}_{XI}$&$|000\rangle+|011\rangle+|121\rangle+|102\rangle$   & $J(\PP^5\times\PP^2,\sigma_2(\PP^1\times\underline{\PP^2}\times\PP^2)\times\PP^2)$ &$13$\\
$\overline{\mathcal{O}}_{X}$&$|000\rangle+|011\rangle+|102\rangle$& $J(X,\sigma_2(\PP^1\times\underline{\PP^2}\times\PP^2)\times\PP^2))$ &$12$\\
$\overline{\mathcal{O}}_{IX}$&$|000\rangle+|011\rangle+|022\rangle$& $\PP^1\times\sigma_3(\PP^2\times\PP^2)\simeq \PP^1\times\PP^8$  &$9$\\
$\overline{\mathcal{O}}_{VIII}$&$|000\rangle+|011\rangle+|110\rangle+|121\rangle$& $\sigma_2(\PP^5\times\PP^2)$  &$13$\\
$\overline{\mathcal{O}}_{VII}$&$|000\rangle+|011\rangle+|120\rangle$& $J(X,\PP^5\times\PP^2)$  &$12$\\
$\overline{\mathcal{O}}_{VI}$&$|000\rangle+|111\rangle$& $\sigma_2(X)$  &$11$\\
$\overline{\mathcal{O}}_{V}$&$|000\rangle+|011\rangle+|101\rangle$& $\tau(X)$  &$10$\\
$\overline{\mathcal{O}}_{IV}$&$|000\rangle+|011\rangle$& $\PP^1\times\sigma_2(\PP^2\times\PP^2)$  &$8$\\
$\overline{\mathcal{O}}_{III}$&$|000\rangle+|101\rangle$& $\sigma_2(\PP^1\times\underline{\PP^2}\times\PP^2)\times\PP^2$  &$7$\\
$\overline{\mathcal{O}}_{II}$&$|000\rangle+|110\rangle$& $\sigma_2(\PP^1\times\PP^2)\times\PP^2\simeq \PP^5\times\PP^2$  &$7$\\
$\overline{\mathcal{O}}_{I}$ &$|000\rangle$& $X=\PP^1\times\PP^2\times\PP^2$  &$5$
 \end{tabular}
\caption{Identification of orbit closures and varieties for $2\times 3\times 3$ quatum system}\label{table233}

\end{center}
\end{table}
 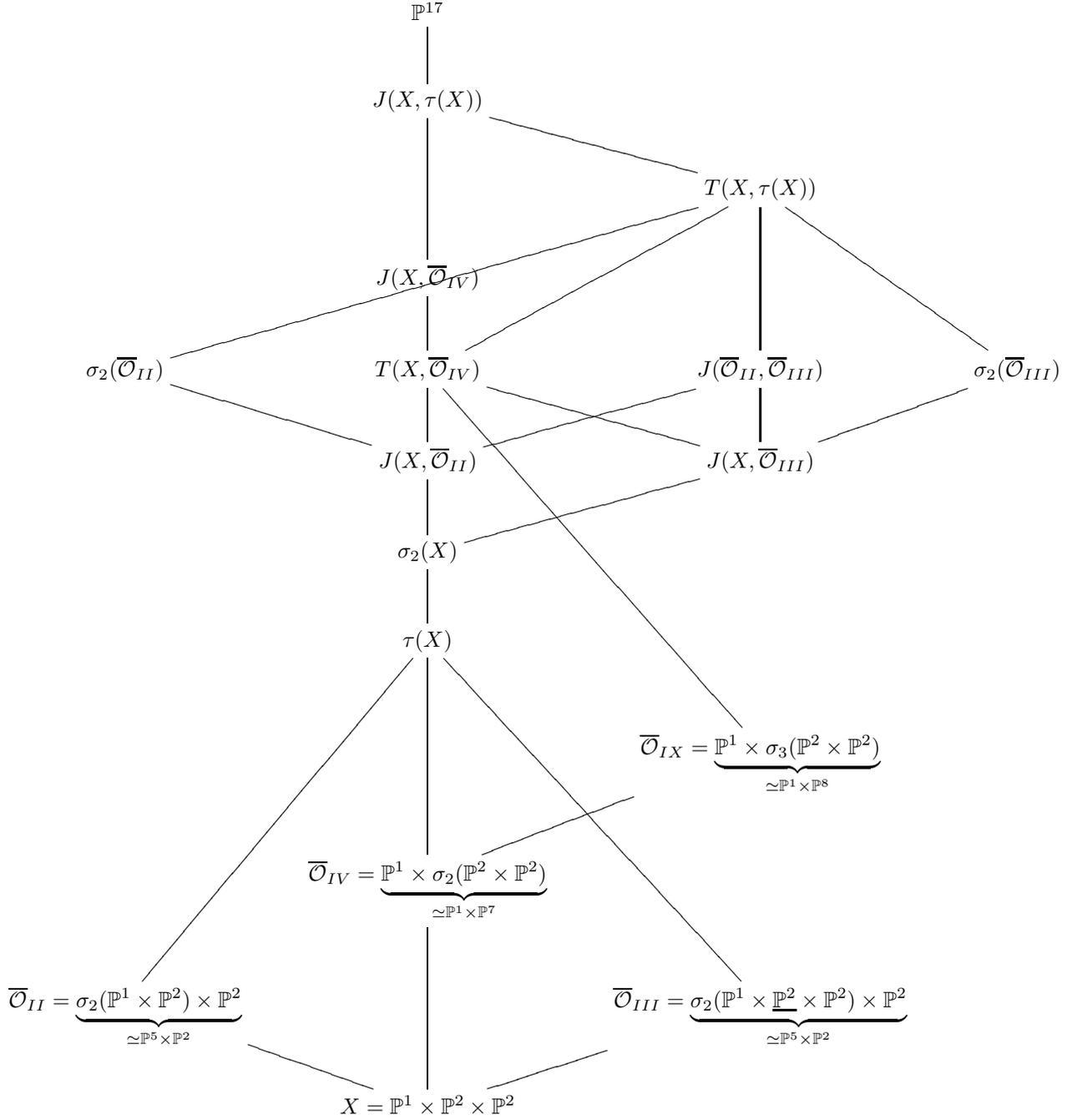
\begin{figure}[!h]
  \[\xymatrix{ & \PP^{17} & &\\
               & J(X,\tau(X))\incl[u] &  &\\
                 &         & T(X,\tau(X))\incl[ul]& \\
              & J(X,\overline{\mathcal{O}}_{IV}) \incl[uu] & &\\
\sigma_2(\overline{\mathcal{O}}_{II}) \incl[uurr] & T(X,\overline{\mathcal{O}}_{IV})\incl[u]\incl[uur] & J(\overline{\mathcal{O}}_{II},\overline{\mathcal{O}}_{III}) \incl[uu]&\sigma_2(\overline{\mathcal{O}}_{III})\incl[uul]\\
 & J(X,\overline{\mathcal{O}}_{II})\incl[ul]\incl[u]\incl[ur] &  J(X,\overline{\mathcal{O}}_{III})\incl[ul]\incl[u]\incl[ur]\\
& \sigma_2(X) \incl[u] \incl[ur]& &\\
& \tau(X) \incl[u] & &\\
&                  &\overline{\mathcal{O}}_{IX}=\underbrace{\PP^1\times\sigma_3(\PP^2\times\PP^2)}_{\simeq \PP^1\times\PP^8} \incl[uuuul]&\\
& \overline{\mathcal{O}}_{IV}=\underbrace{\PP^1\times\sigma_2(\PP^2\times \PP^2)}_{\simeq \PP^1\times\PP^7}\incl[uu]\incl[ur]& \\
\overline{\mathcal{O}}_{II}=\underbrace{\sigma_2(\PP^1\times \PP^2)\times\PP^2}_{\simeq \PP^5\times\PP^2}\incl[uuur] &  &\overline{\mathcal{O}}_{III}=\underbrace{\sigma_2(\PP^1\times\underline{\PP^2}\times\PP^2)\times\PP^2}_{\simeq \PP^5\times\PP^2} \incl[uuul]&\\
& X=\PP^1\times \PP^2 \times \PP^2 \incl[uu] \incl[ru] \incl[ul] & &\\
  }\]
\caption{Stratification of the ambient space for the $2\times 3\times 3$ quantum system}\label{233onion}
 \end{figure}
\end{theorem}

\begin{rem}\rm
 The geometric inclusions between the varieties of Tables \ref{table222}, \ref{table223}, \ref{table22n} and \ref{table233} are represented in  Figures 
\ref{222onion}, \ref{223onion}, \ref{22nonion} and \ref{233onion}. Those inclusions can be deduced from the relations between the representatives 
of each orbit, 
e.g. in Table \ref{table233} it is clear that  $\overline{\mathcal{O}}_{IX}$ is a subvariety of  $\overline{\mathcal{O}}_{XIII}$. But it could also 
be deduced from geometric considerations, e.g. the tangential varieties are included in the secant varieties (i.e. $\overline{\mathcal{O}}_V \subset \overline{\mathcal{O}}_{VI}$). 
The geometry of the auxiliary varieties reflects how inequivalent entangled states are partially ordered under local actions.
This was emphasized in \cite{My,My2} and it becomes very natural in our description by auxiliary varieties.
\end{rem}

\begin{rem}\rm
As we point it out in Section \ref{algo} a lot of classifications of entanglement in quantum systems can be deduced from older results from representation theory and
invariant theory. 
For instance for the $2\times 3\times 3$ system, the classification was first established in QIT context in \cite{chen} but can 
be deduced from \cite{Pav}.
It is interesting to notice that the classifications of tripartite entangled states of \cite{chen,My,My3,My2} can be obtained  
from  a classification theorem on 
trilinear forms of type $2\times n\times n$ proved by Camille Jordan in 1907 \cite{jordan}.
\end{rem}

\subsection*{Proofs of Theorem \ref{identification} and Theorem \ref{identification2}}\label{proof}
We now prove the identifications of the orbit closures with algebraic varieties constructed from the unique closed orbit under the action of $G=SL_p\times SL_q\times SL_r$ 
on $\PP(\CC^p\otimes \CC^q\otimes \CC^r)$ for \[(p,q,r)=\left\{\begin{array}{ll}
2\times 2\times 2 & \text{Table }\ref{table222}\\
2\times 2\times 3 & \text{Table }\ref{table223}\\
                                                          2\times 2\times(n+1), n\geq 3 & \text{Table }\ref{table22n}\\
                                                          2\times 3\times 3 &\text{Table }\ref{table233}
                                                         \end{array}\right.\]
The normal forms (representatives) of the orbits under the action of $G$ are given in \cite{Pav} and there is a finite number of them. We express those normal
forms in the tables with the braket notation by the convention of the introduction $|ijk\rangle=e_i\otimes e_j\otimes e_k$ where $(e_t)_{0\leq t\leq s-1}$ stands for a
 basis of $\CC^{s}$.
The first step in our proof is to calculate the dimension of each orbit closure. It can be done directly using the following two facts:
\begin{itemize}
 \item Let $\overline{\mathcal{O}_x}=\PP(\overline{G.x})\subset \PP(V)$ be the closure of a $G$-orbit with representative $x$, for $G$ a semi-simple Lie group with Lie algebra $\gG$. 
Then $\widehat{T}_x\mathcal{O}=\gG.x$.
The notation $\gG.x$ represents the action of $\gG$ on the representation $V$ (\cite{F-H}). Moreover if $x=u+v$ one has $\widehat{T}_x\mathcal{O}=\gG.x=\gG.u+\gG.v$.
\item Consider the Segre product $X=\PP(G.(e_1\otimes\dots\otimes e_m))=Y_1\times\dots\times Y_m\subset \PP(V_1\otimes \dots\otimes V_m)$ with $G$ a semi-simple Lie group, then we have 
$\widehat{T}_{e_1\otimes\dots\otimes e_m}X=\widehat{T}_{e_1}Y_1\otimes e_2\otimes\dots\otimes e_m+e_1\otimes \widehat{T}_{e_2}Y_2\otimes \dots\otimes e_m+\dots+e_1\otimes e_2\otimes\dots\otimes \widehat{T}_{e_m}Y_m$. 
In particular for $X=\PP^{p-1}\times \PP^{q-1}\times\PP^{r-1}$ and $e\otimes f\otimes g \in \widehat{X}$, we have $\widehat{T}_{e\otimes f\otimes g}=\CC^p\otimes f\otimes g+e\otimes\CC^q\otimes g+e\otimes f\otimes \CC^r$.
\end{itemize}
 
\noindent Those remarks allow us to calculate the dimension for any orbit from the normal form. For instance for the orbit closure $\overline{\mathcal{O}}_{VI}$ we get:
 
\[\widehat{T}_{|000\rangle+|111\rangle}G.(|000\rangle+|111\rangle)=\widehat{T}_{|000\rangle}G.|000\rangle+\widehat{T}_{|111\rangle}G.|111\rangle\]

\begin{equation}\label{calc}
=\underbrace{\CC^p\otimes e_0\otimes e_0+e_0\otimes \CC^q\otimes e_0+e_0\otimes
e_0\otimes \CC^r+\CC^p\otimes e_1\otimes e_1+e_1\otimes \CC^q\otimes e_1+e_1\otimes
e_1\otimes \CC^r}_{\text{dim}=2(p+q+r)-4}
\end{equation}
After projectivization $\text{dim}(\overline{G.(|000\rangle+|111\rangle)})=2(p+q+r)-5$. 
That orbit is clearly a subvariety of $\sigma_2(\PP^{p-1}\times\PP^{q-1}\times \PP^{r-1})$ and $\text{dim}(\sigma_2(X))\leq 2(p-1+q-1+r-1)+1$. 
The variety $\sigma_2(X)$ is irreducible because $X$ is. Thus we have 
$\overline{G.(|000\rangle+|111\rangle)}\subset\sigma_2(\PP^{p-1}\times\PP^{q-1}\times \PP^{r-1})$ with equality of dimensions which proves the equality 
$\overline{G.(|000\rangle+|111\rangle)}=\sigma_2(\PP^{p-1}\times\PP^{q-1}\times \PP^{r-1})$.
The secant variety is of maximal dimension then one deduces from Proposition \ref{secant-tangent} that the tangential variety $\tau(\PP^{p-1}\times\PP^{q-1}\times \PP^{r-1})$
is of dimension $2(p+q+r)-6$ and we identify it with the orbit $\overline{\mathcal{O}}_{V}$ which is the only one of dimension $2(p+q+r)-6$.
For the orbits $\overline{\mathcal{O}}_{II}$ and $\overline{\mathcal{O}}_{IV}$ one 
notices that $|000\rangle+|011\rangle=e_0\otimes(e_{0}\otimes e_0+e_1\otimes e_1)$ and $|000\rangle+|110\rangle=(e_0\otimes e_0+e_1\otimes e_1)\otimes e_0$  which allows us 
to identify those orbit closures with $\sigma_2(\PP^{p-1}\otimes\PP^{q-1})\times\PP^{r-1}$ and $\PP^{p-1}\times \sigma_2(\PP^{q-1}\times\PP^{r-1})$ after a dimension count 
(again the orbit is clearly a
subvariety of the $\{1\}$-subsecant or the $\{3\}$-subsecant and has the same dimension). The orbit closure $\PP(\overline{G.(|010\rangle+|111\rangle)})$ is 
the orbit of the line $\PP^1_{e_{0}\otimes e_1\otimes e_0,e_1\otimes e_1\otimes e_1}$ defined by the $\{2\}$-pair of points $(e_{0}\otimes e_1\otimes e_0,e_1\otimes e_1\otimes e_1)$. 
The corresponding orbit closure is the $\{2\}$-subsecant variety $\sigma_2(\PP^1\times \underline{\PP^1}\times\PP^1)\times\PP^1$. 
This completes the identifications of Table \ref{table222}. 

\noindent For Tables \ref{table223} and \ref{table22n} we observe, after a dimension calculation similar to (\ref{calc}), that the
 dimension of $\overline{\mathcal{O}}_{VII}$ is equal to $3n+4$ ($n=2$ for Table \ref{table223} and $n\geq 3$ for Table \ref{table22n}). 
Moreover it is clear from the representatives that $J(X,\overline{\mathcal{O}}_{IV})\supset \overline{\mathcal{O}}_{VII}$. 
That last assertion comes from
\[x_{VII}=\underbrace{|000\rangle+|011\rangle}_{\in \overline{\mathcal{O}}_{IV}}+\underbrace{|102\rangle}_{\in X} \in J(X,\overline{\mathcal{O}}_{IV})\]
We also know that $\text{dim}(J(X,\overline{\mathcal{O}}_{IV}))\leq \text{dim}(X)+\text{dim}(\overline{\mathcal{O}}_{IV})+1=3n+5$. 
Thus we have  $\overline{\mathcal{O}}_{VII}\subset J(X,\overline{\mathcal{O}}_{IV})$ and $3n+4=\text{dim}(\overline{\mathcal{O}}_{VII})\leq \text{dim}(J(X,\overline{\mathcal{O}}_{IV}))\leq 3n+5$.
We now prove with  Teracini's Lemma that $J(X,\overline{\mathcal{O}}_{IV})$ is in fact of dimension $3n+4$.
 Let $x=e\otimes f\otimes g+h\otimes(m\otimes n+p\otimes q)$ be a smooth point of 
$J(X,\overline{\mathcal{O}}_{IV})$.  Teracini's Lemma says that the tangent space of the join is given by
 $\widehat{T}_x J(X,\overline{\mathcal{O}}_{IV})=\widehat{T}_{e\otimes f\otimes g} X+\widehat{T}_{h\otimes (m\otimes n+p\otimes q)} \overline{\mathcal{O}}_{IV}
$ with $\widehat{T}_{e\otimes f\otimes g} X=\CC^2\otimes f\otimes g+e\otimes\CC^2\otimes g+e\otimes f\otimes \CC^{n+1}$ 
and $\widehat{T}_{h\otimes(m\otimes n+p\otimes q)} \overline{\mathcal{O}}_{IV}=\CC^2\otimes(m\otimes n+p\otimes q)+h\otimes \CC^{2n+2}$. 
When we look at the intersection of the tangent spaces we have $h\otimes f\otimes g\in \widehat{T}_{e\otimes f\otimes g} X\cap \widehat{T}_{h\otimes(m\otimes n+p\otimes q)} \overline{\mathcal{O}}_{IV}$, thus the intersection does not reduce to $\{0\}$ and therefore the join 
is not of maximal dimension, i.e. $\text{dim}(J(X,\overline{\mathcal{O}}_{IV}))\leq 3n+4$. We conclude that $\text{dim}(J(X,\overline{\mathcal{O}}_{IV}))= 3n+4$ and corresponds to $\overline{\mathcal{O}}_{VII}$.
The orbit $\overline{\mathcal{O}}_{VIII}$ is of dimension $11$ for Table \ref{table223} and corresponds to the ambient space $\PP^{11}$. The orbit $\overline{\mathcal{O}}_{VIII}$ is of dimension 
$3n+5$ for Table \ref{table22n} but it is known, \cite{Lan2}, that $\sigma_3(\PP^{p-1}\times \PP^{q-1}\times \PP^{r-1})$ is of dimension $3p+3q+3r-7$ and therefore 
we can state that $\overline{\mathcal{O}}_{VIII}=\sigma_3(\PP^1\times\PP^1\times\PP^n)$.

\noindent The last identifications concern Table \ref{table233}:
with Teracini's Lemma one shows that $J(X,\tau(X))$ has the expected dimension, i.e. $16$. 
Then the orbit closure $\overline{\mathcal{O}}_{XVI}$, which is the unique orbit of dimension $16$, corresponds to $J(X,\tau(X))$ 
(indeed we have $J(X,\tau(X))=T(X,\sigma_2(X))$ because here $\sigma_3(X)$ fills the ambient space).
 One deduces there exists an orbit closure
of dimension $15$ which corresponds to $T(X,\tau(X))$. This has to be the orbit closure  $\overline{\mathcal{O}}_{XV}$ (the only orbit of dimension $15$).
It is clear from the normal form that $\overline{\mathcal{O}}_{XIV}$ is included in $J(X,\PP^1\times\sigma_2(\PP^2\times\PP^2))$. 
But
the expected dimension of $J(X,\PP^1\times\sigma_2(\PP^2\times\PP^2))$ is $14$ which is the dimension of $\overline{\mathcal{O}}_{XIV}$.
 Thus we conclude to the equality between $\overline{\mathcal{O}}_{XIV}$ and  $J(X,\PP^1\times\sigma_2(\PP^2\times\PP^2))$.
As $J(X,\PP^1\times\sigma_2(\PP^2\times\PP^2))$ has the expected dimension one knows there exists a variety of dimension $13$ corresponding to 
$T(X,\PP^1\times\sigma_2(\PP^2\times\PP^2))$. But according to \cite{Pav} there is only one orbit of dimension $13$ in $\overline{\mathcal{O}}_{XIV}$ and 
that is 
${\mathcal{O}}_{XIII}$. Thus the orbit closure $\overline{\mathcal{O}}_{XIII}$ corresponds to  $T(X,\PP^1\times\sigma_2(\PP^2\times\PP^2))$.
The varieties $\sigma_2(\PP^1\times\PP^2)\times\PP^2$ and $\sigma_2(\PP^1\times\underline{\PP^2}\times\PP^2)\times\PP^2$ 
are isomorphic to $\PP^5\times\PP^2$ and
$\sigma_2(\PP^5\times\PP^2)$ is of dimension $13$ (that's the projectivization of the set of rank at least 
$2$ matrices in the projectivization of the space of
$6\times 3$ matrices). But there are three others orbits of dimension $13$ which are orbits $\mathcal{O}_{VIII}$, $\mathcal{O}_{XI}$ and $\mathcal{O}_{XII}$.
 From the normal forms we can affirm
that  orbit ${\mathcal{O}}_{XII}$ is contained 
in $\sigma_2(\sigma_2(\PP^1\times\underline{\PP^2}\times\PP^2)\times\PP^2)$ and the orbit $\mathcal{O}_{VIII}$ is contained in $\sigma_2(\sigma_2(\PP^1\times\PP^2)\times\PP^2)=\sigma_2(\PP^5\times\PP^2)$
because
\[x_{XII}=|000\rangle+|011\rangle+|101\rangle+|112\rangle=\underbrace{|000\rangle+|101\rangle}_{\in\sigma_2(\PP^1\times\underline{\PP^2}\times\PP^2)\times\PP^2} +\underbrace{|011\rangle+|112\rangle}_{\in\sigma_2(\PP^1\times\underline{\PP^2}\times\PP^2)\times\PP^2}\]

\[x_{VIII}=|000\rangle+|011\rangle+|110\rangle+|121\rangle=\underbrace{|000\rangle+|011\rangle}_{\in\sigma_2(\PP^1\times\PP^2)\times\PP^2} +\underbrace{|110\rangle+|121\rangle}_{\in\sigma_2(\PP^1\times\times\PP^2)\times\PP^2}\]

\noindent That leads to the equalities $\overline{\mathcal{O}}_{XII}=\sigma_2(\sigma_2(\PP^1\times\underline{\PP^2}\times\PP^2)\times\PP^2)$ and 
$\overline{\mathcal{O}}_{VIII}=\sigma_2(\PP^5\times\PP^2)$.
The last orbit closure of dimension $13$, namely $\overline{\mathcal{O}}_{XI}$, is a subvariety of $J(\PP^5\times\PP^2,\sigma_2(\PP^1\times\underline{\PP^2}\times\PP^2)\times\PP^2)$:
\[x_{XI}=\underbrace{|000\rangle+|102\rangle}_{\in \sigma(\PP^1\times\underline{\PP^2}\times\PP^2)\times\PP^2}+\underbrace{|011\rangle+|121\rangle}_{\in \PP^5\times\PP^2}\]

\noindent By Teracini's Lemma one obtains that  $J(\PP^5\times\PP^2,\sigma_2(\PP^1\times\underline{\PP^2}\times\PP^2)\times\PP^2)$ is of dimension less than $13$. 
Indeed let $u=(a\otimes b+c\otimes d)\otimes e\in \PP^5\times \PP^2$, then $\widehat{T}_u\PP^5\times\PP^2=\CC^6\otimes e+(a\otimes b+c\otimes d)\otimes \CC^3$ 
and $v=f\otimes g\otimes h+k\otimes g\otimes l\in \sigma_2(\PP^1\times\underline{\PP^2}\times\PP^2)\times\PP^2$, then 
$\widehat{T}_v\sigma_2(\PP^1\times\underline{\PP^2}\times\PP^2)\times\PP^2=\CC^2\otimes g\otimes\CC^3+(f\otimes \underline{g}\otimes h+k\otimes \underline{g}\otimes l)\otimes\CC^3$.
Thus $(\widehat{T}_u\PP^5\times\PP^2)\cap (\widehat{T}_v\sigma_2(\PP^1\times\underline{\PP^2}\times\PP^2)\times\PP^2)\supset \CC^2\otimes g\otimes e$, i.e. the dimension 
of the join variety is at most $15-2=13$ and therefore we have $\overline{\mathcal{O}}_{{XI}}=J(\PP^5\times\PP^2,\sigma_2(\PP^1\times\underline{\PP^2}\times\PP^2)\times\PP^2)$.

\noindent The same argument holds for orbits $\mathcal{O}_{VII}$ and $\mathcal{O}_{X}$. One first shows that $\overline{\mathcal{O}}_{VII}\subset J(X,\PP^5\times\PP^2)$ and $\overline{\mathcal{O}}_{\text{X}}\subset J(X,\sigma_2(\PP^1\times\underline{\PP^2}\times\PP^2)\times\PP^2)$:

 \[x_{VII}=\underbrace{|000\rangle+|120\rangle}_{\in \sigma_2(\PP^1\times\PP^2)\times\PP^2}+\underbrace{|011\rangle}_{\in X}\]

 \[x_{X}=\underbrace{|000\rangle+|102\rangle}_{\in \sigma (\PP^1\times\underline{\PP^2}\times\PP^2)\times\PP^2}+\underbrace{|011\rangle}_{\in X}\]

\noindent Then  Teracini's Lemma allows us to prove that the join varieties are of dimension less than $12$ and we conclude to the equality.

\noindent The last orbit to identify is $\mathcal{O}_{{IX}}$ which is of dimension $9$. From the normal form this orbit is clearly included in $\PP^1\times\sigma_3(\PP^2\times\PP^2)=\PP^1\times\PP^8$ and 
the identification follows because of the equality of dimensions. $\Box$

\section{Back to Miyake's geometric description by dual varieties}\label{dual}

 Since $G$ acts with a finite number of orbits on $\PP(\mathcal{H})$, the orbit structure of 
$\PP(\mathcal{H}^*)$ is identical to the orbit structure of $\PP(\mathcal{H})$. In this section we identify the orbit closures in 
$\PP(\mathcal{H}^*)$ with duals of varieties of the stratification given by Theorem \ref{identification}. We then recover Miyake's geometric
descriptions of entangled states by dual varieties for the $2\times 2\times 2$ and $2\times 2\times(n+1)$ quantum systems.
\subsection*{The dual variety and its singular locus}
Let  $X\subset \PP(V)$ be a projective variety and let $\tilde{T}_x
  X$ denote the embedded tangent space of $X$ at $x$, a smooth point of $X$. Define the {\em dual variety}  $X^*$ by

\[X^*=\overline{\{H \in \PP(V^*) | \ \exists \ x \in X_{smooth} \text{ such that} \  \tilde{T}_x X\subset H\}}\subset\PP(V^*)\]
The biduality theorem $(X^*)^*=X$, true in charestic zero, implies that the original variety can be reconstructed from its dual variety. The dual varieties have been studied 
intensively by algebraic geometers (\cite{Tev}).
In the case of the variety  $X=\PP^n\times\PP^n\subset \PP^{(n+1)^2-1}$ (projectivization of rank one matrices) it is well known that 
the dual variety can be identified
 with the variety of  matrices of rank at most $n$i. Thus, up to multiplication by a nonzero scalar the equation defining $X^*$ is the determinant. This leads to
a higher dimensional generalization of the determinant, called hyperdeterminant, which was first introduced by Cayley and rediscovered by Gelfand, Kapranov and Zelevinsky \cite{GKZ}.
The hyperdeterminant in the sense of \cite{GKZ} is the defining equation of the dual of $X=\PP^{n_1}\times\dots\times \PP^{n_k}$ when $X^*$ is a hypersurface.

\noindent In \cite{My} A. Miyake uses this notion of hyperdeterminant to classify multipartite entangled states for $2\times 2\times n$ quantum systems. 
In the bipartite case, the dual of the set of separable states, $\PP^n\times\PP^n\subset \PP^{(n+1)^2-1}$, is isomorphic to the 
the projectivization of the set of  matrices of rank less than $n$. The generic entangled state, in the bipartite case, corresponds to matrices of maximal rank and
  therefore corresponds to $\PP^{(n+1)^2-1}\backslash X^*$. 
Following the analogy with the 2-dimensional case Miyake proposes to use the stratification of $\PP(V^*)$ by $X^*$ and its subvarieties to distinguish the states of entanglement.
The variety of separable states $X$ being SLOCC invariant so is $X^*$ 
and its singular locus. The dual variety and its singularities induce a filtration of the (dual) ambient space:

\[X_{sing}^*\subset X^*\subset \PP(V^*)\]

\noindent In order to explain what the singular locus of the dual variety is, we need to look at 
the tangent hyperplanes of the variety $X$.
When $X^*$ is a hypersurface, a smooth hyperplane $H\in X^*$ is a hyperplane 
tangent to $X$ with a unique singular point which is a nondegenerate quadric. In other words the restriction to $X$ of the linear form which defines $H$
is a quadric of full rank (a hypersurface with a $A_1$ singular point).
Thus when $X^*$ is a hypersuface there are two ways for a hyperplane to not be a smooth point, either by having more than one point of tangency or by defining
a degenerate quadric \cite{FH,WZ}.

\begin{def }
Let $X\subset \PP(V)$ and $X^*$ be its dual variety, which we assume to be a hypersurface. We define $X^*_{{node}}$, the node 
component of $X^*$ to be the set of hyperplanes having more than one point of tangency :
\[X^*_{node}=\{H\in X^*, \exists (x,y)\in X\times X, x\neq y, \tilde{T}_x X\subset H,\tilde{T}_yX\subset H\}\]
We define the cusp component $X^*_{\text{cusp}}$, to be the set of hyperplanes defining a singular hyperplane section with degenerate quadratic part :
\[X^*_{cups}=\{H\in X^*, \exists x\in X, \tilde{T}_x X\subset H, (X\cap H,x)\not\sim A_1\}\]
\end{def }

\noindent Following \cite{GKZ} we can decompose $X^*_{node}$ into irreducible components :

\begin{def } Let $X=Y^{n_1}\times\dots Y^{n_m}\subset \PP^{(n_1+1)\dots (n_m+1)-1}$ be the Segre product of $m$ nondegenerate varieties and $J=\{j_1,\dots,j_k\}\subset \{1,\dots,m\}$
 \[X^* _{\text{node}}(J)=\overline{\{H\in \PP(V^*), \exists (x,y) \text{ a } J-\text{pair of point of } X\times X, \tilde{T}_xX\subset H,\tilde{T}_xX\subset H\}}\]
\end{def }

\noindent We now prove a proposition which describes for Segre products the node components in terms of the dual of the subsecant varieties.

\begin{prop}\label{dual-Jsec}
Let $X=Y_1\times \dots\times Y_m\subset \PP(V)$ be the Segre product of $m$ nondegenerate varieties and let $X^*$ be 
its dual variety.
 The $J$-node component is the dual of the $J$-subsecant variety, i.e. for $J=\{j_1,\dots,j_k\}\subset\{1,\dots,m\}$:
\[X^*_{\text{node}}(J)=(\sigma_2(Y_1\times\dots \times\underline{Y}_{j_1}\times \dots\times \underline{Y}_{j_k}\times \dots\times Y_m)\times Y_{j_1}\times Y_{j_2}\times\dots\times Y_{j_k})^*\]

\end{prop}

\proof It is a consequence of  Teracini's Lemma. Let us denote by $Z$ the $J$-subsecant variety 
$Z=\sigma_2(Y_1\times\dots \times\underline{Y}_{j_1}\times \dots\times \underline{Y}_{j_k}\times \dots\times Y_m)\times Y_{j_1}\times Y_{j_2}\times\dots\times Y_{j_k}$
 and $z=x+y\in Z$ a general point of $Z$. By definition $(x,y)$ is a $J$-pair of point. According to the Teracini's Lemma we have $\widehat{T}_x X+\widehat{T}_yX= \widehat{T}_z Z$. 
Thus if 
$H\in Z^*$ is tangent to $Z$ at  $z$ it means $\widehat{H}\supset \widehat{T}_x X$ and $\widehat{H}\supset \widehat{T}_y X$ i.e. 
$H\in X^*_{\text{node}}(J)$ because $(x,y)$ is a $J$-pair of point. On the other hand if 
$H\in X^*_{node}(J)$ then there exists a $J$-pair of point $(x,y)$ such that $\widehat{H}\supset \widehat{T}_x X$ and $\widehat{H}\supset \widehat{T}_yX$ i.e. $H$ is tangent
 to $Z$ at the point $z=x+y$, i.e. $H\in Z^*$. $\Box$.

\begin{rem}\rm
This proposition shows in particular that $X^*_{node}(\emptyset)=\sigma_2(X)^*$. This equality is used
 in \cite{FH} to study the dimension of the singular locus of the duals of Grassmannians.
\end{rem}

\subsection*{Stratification by the dual variety and its singular locus}

We now recover Miyake's classifications \cite{My,My2} by dual varieties by establishing the isomorphisms between 
the varieties of Theorem \ref{identification}, and the varieties in the dual space. 
Morevover the components of the singular locus of the dual variety are described in terms of the duals of auxiliary varieties.
\begin{theorem}\label{dualorbit}
For $2\times 2\times(n+1)$ ($n\geq 1$) quantum systems the duality 
between the orbit closures are given in Tables \ref{table222d}, \ref{table223d} and  \ref{table22nd}
\begin{table}[!h]
\begin{center}
\begin{tabular}{c|c}
\hline
Orbits & Varieties \\
\hline

$\overline{\mathcal{O}}_{V}\simeq \overline{\mathcal{O}}_{I} ^*$& $\tau(\PP^1\times\PP^1\times\PP^1)\simeq(\PP^1\times\PP^1\times\PP^1)^*$ \\
$\overline{\mathcal{O}}_{IV}\simeq \overline{\mathcal{O}}_{IV}^*$  & $(\sigma_2(\PP^1\times\PP^1)\times \PP^1)\simeq(\sigma_2(\PP^1\times\PP^1)\times \PP^1)^*=X^*_{node}(\{3\})$ \\ 
$\overline{\mathcal{O}}_{III}\simeq \overline{\mathcal{O}}_{III}^*$  &$ (\PP^1\times\sigma_2(\PP^1\times \PP^1))\simeq(\PP^1\times\sigma_2(\PP^1\times \PP^1))^*=X^*_{node}(\{1\})$ \\
$\overline{\mathcal{O}}_{II}\simeq \overline{\mathcal{O}}_{II}^*$ & $(\sigma_2(\PP^1\times\underline{\PP^1}\times \PP^1)\times \PP^1)\simeq(\sigma_2(\PP^1\times\underline{\PP^1}\times \PP^1)\times \PP^1)^*=X^*_{node}(\{2\})$  \\
\end{tabular}
\caption{Duality between  orbit closures for the $2\times 2\times 2$ quantum system}\label{table222d}
\end{center}
\end{table}

\begin{table}[!h]
\begin{center}
\begin{tabular}{c|c}
\hline
Orbits  & Varieties \\
\hline
$\overline{\mathcal{O}}_{VII}\simeq \overline{\mathcal{O}}_I ^*$& $J(\PP^1\times\PP^1\times\PP^2,\PP^1\times\sigma(\PP^1\times\PP^2))\simeq (\PP^1\times\PP^1\times\PP^2)^*$ \\
$\overline{\mathcal{O}}_{VI}\simeq \mathcal{\overline{O}}_{II}^*$ & $\sigma_2(\PP^1\times\PP^1\times\PP^2)\simeq (\sigma_2(\PP^1\times\PP^1)\times\PP^2)^*\simeq (\PP^3\times\PP^2)^*=X^*_{node}(\{3\})$  \\
$\overline{\mathcal{O}}_{V}\simeq \overline{\mathcal{O}}_V ^*$& $\tau(\PP^1\times\PP^1\times\PP^2)\simeq \tau(\PP^1\times\PP^1\times\PP^2)^*$  \\
$\overline{\mathcal{O}}_{IV}\simeq \overline{\mathcal{O}}_{IV} ^*$& $\PP^1\times \sigma_2(\PP^1\times\PP^2)\simeq(\PP^1\times\sigma_2(\PP^1\times \PP^2))^*=X^*_{node}(\{1\})$\\
$\overline{\mathcal{O}}_{III}\simeq\overline{\mathcal{O}}_{III}^*$& $\sigma_2(\PP^1\times \underline{\PP^1}\times \PP^2)\times \PP^1\simeq(\sigma_2(\PP^1\times \underline{\PP^1}\times \PP^2)\times \PP^1)^*=X_{node} ^*(\{2\})$ \\
\end{tabular}
\caption{Duality between the orbit closures for the $2\times 2\times 3$ quatum system}\label{table223d}
\end{center}
\end{table}

\begin{table}[!h]
\begin{center}
\begin{tabular}{c|c}
\hline
Orbits& Varieties \\
\hline

$\overline{\mathcal{O}}_{VIII}\simeq \overline{\mathcal{O}}_{II}^*$ & $\sigma_3(\PP^1\times\PP^1\times\PP^n)\simeq (\sigma_2(\PP^1\times \PP^1)\times\PP^n)^*\simeq(\PP^3\times\PP^n)^*$   \\
$\overline{\mathcal{O}}_{VII}\simeq \overline{\mathcal{O}}_{I}^*$& $J(\PP^1\times\PP^1\times\PP^n,\PP^1\times\sigma_2(\PP^1\times\PP^n))\simeq (\PP^1\times\PP^1\times\PP^n)^*$\\
$\overline{\mathcal{O}}_{VI}\simeq \overline{\mathcal{O}}_{VI}^*$& $\sigma_2(\PP^1\times\PP^1\times\PP^n)\simeq \sigma_2(\PP^1\times\PP^1\times\PP^n)^*$   \\
$\overline{\mathcal{O}}_{V}\simeq \overline{\mathcal{O}}_{V}^*$& $\tau(\PP^1\times\PP^1\times\PP^n)\simeq \tau(\PP^1\times\PP^1\times\PP^n)^*$\\
$\overline{\mathcal{O}}_{IV}\simeq \overline{\mathcal{O}}_{IV}^*$& $(\PP^1\times \sigma_2(\PP^1\times \PP^n)\simeq \PP^1\times\PP^{2n+1}\simeq (\PP^1\times \PP^{2n+1})^*\simeq \PP^1\times(\sigma_2(\PP^1\times \PP^n))^*$\\
$\overline{\mathcal{O}}_{III}\simeq \overline{\mathcal{O}}_{III}^* $ & $\sigma_2(\PP^1\times \underline{\PP^1}\times \PP^n)\times \PP^1\simeq (\sigma_2(\PP^1\times \underline{\PP^1}\times \PP^n)\times \PP^1)^*$ 
\end{tabular}
\caption{Duality between the orbit closures for the $2\times2\times(n+1)$ quantum systems, $n\geq 3$}\label{table22nd}
\end{center}
\end{table}

\end{theorem}

\proof Most of the identifications follow from calculation of the dimension of the dual of each variety of Theorem \ref{identification}.

\begin{enumerate}
 \item Table \ref{table222d}: it is well known (\cite{KM}) the dual of $\PP^1\times\PP^1\times\PP^1$ is a hypersuface and 
therefore corresponds to the closure of
the unique orbit of dimension $6$, i.e. $(\PP^1\times\PP^1\times\PP^1)^*\simeq \tau(\PP^1\times\PP^1\times\PP^1)$.

 The $\{j\}$-subsecant varieties  are isomorphic to $\PP^3\times\PP^1$. 
But $\PP^3\times\PP^1$ is the projectivization of rank one $4\times 2$ matrices. The set of rank one $4\times 2$ matrices is equal to the set of degenerate
$4\times 2$ matrices (the generic $4\times 2$ matrices are of rank $2$ and the degenerate ones of rank strictly less than $2$). 
It means $\PP^3\times\PP^1$ is self dual, i.e. $\PP^3\times\PP^1\simeq (\PP^3\times\PP^1)^*$ 
and therefore so are the $\{j\}$-subsecant varieties.
\item Table \ref{table223d}: here again the dual of $\PP^1\times\PP^1\times\PP^2$ is a hypersurface and thus according to Theorem \ref{identification} one gets
$(\PP^1\times\PP^1\times\PP^2)^*\simeq J(\PP^1\times\PP^1\times\PP^2,\PP^1\times\sigma_2(\PP^1\times\PP^2))$ (the unique hypersurface).
The orbit closure $\overline{\mathcal{O}}_{II}$ is isomorphic to $\PP^3\times \PP^2$. But the projectivization of the 
set of rank one $4\times 3$ matrices is dual to the (projectivization of the ) set of rank at most
$2$ matrices which is of dimension $9$ (after projectivization). Thus we identify $\overline{\mathcal{O}}_{II}^*$ and the unique orbit closure 
of dimension $9$, which is $\overline{\mathcal{O}}_{VI}$. The orbit closures $\overline{\mathcal{O}}_{III}$ and $\overline{\mathcal{O}}_{IV}$ are self-dual because isomorphic to $\PP^1\times\PP^5$.
The remaining variety $ \overline{\mathcal{O}}_{V}$ has to be self-dual.
\item Table \ref{table22nd}: in this table the dual of $\PP^1\times\PP^1\times\PP^n$ ($n\geq 3$) is no longer a hypersurface. 
A dimension count, using Katz's dimension formula (\cite{katz}), shows that $(\PP^1\times\PP^1\times\PP^n)^*$ is of dimension $3n+4$ 
and therefore is isomorphic to $J(\PP^1\times\PP^1\times \PP^n,\PP^1\times\sigma_2(\PP^1\times\PP^n))$ according to Theorem \ref{identification}.
The orbit $\overline{\mathcal{O}}_{II}$ is isomorphic to $\PP^3\times\PP^n$ the projectivization of rank one $4\times (n+1)$ matrices. Its dual variety is
the projectivization of the $4\times (n+1)$ matrices of rank less than $3$. But the projectivization of $4\times(n+1)$ matrices of rank less than $3$ is of dimension 
$3n+5$ and therefore $\overline{\mathcal{O}}_{II}^*\simeq \overline{\mathcal{O}}_{VIII}$. The orbit closures $\overline{\mathcal{O}}_{III}$ and $\overline{\mathcal{O}}_{IV}$
are isomorphic to $\PP^1\times\PP^{2n+1}$ and therefore are self dual. To conclude one calculates the dimension of 
$\sigma_2(\PP^1\times\PP^1\times\PP^n)^*$ using techniques proposed in \cite{FH} where the dimension of $\sigma_2(X)^*$ has been studied. Those calculations lead
to $\text{dim}(\sigma_2(\PP^1\times\PP^1\times\PP^n)^*)=2n+5$ and therefore the secant variety is self-dual. It forces the tangential variety to be also self-dual. $\Box$

 \end{enumerate}

\begin{figure}[!h]
\[\xymatrix{  &  (\PP^7)^*&  \\ &(\PP^1\times\PP^1\times\PP^1)^*\incl[dr]\incl[u] &  \\
      {}\underbrace{(\sigma_2(\PP^1\times \PP^1)\times\PP^1)^*}_{X^*_{node}(3)}\incl[ru]\incl[rd] &{}\underbrace{(\PP^1\times\sigma_2(\PP^1\times \PP^1))^*}_{X^*_{node}(1)}\incl[u] \incl[d]& {}\underbrace{(\sigma_2(\PP^1\times\underline{\PP^1}\times\PP^1)\times\PP^1)^*}_{X^*_{node}(2)} \\
 & \PP^1\times \PP^1 \times \PP^1\simeq \tau(\PP^1\times\PP^1\times\PP^1) ^* \incl[ur] &\\
}  \]
\caption{Stratification by the dual variety and its singular locus}\label{222donion}
\end{figure}
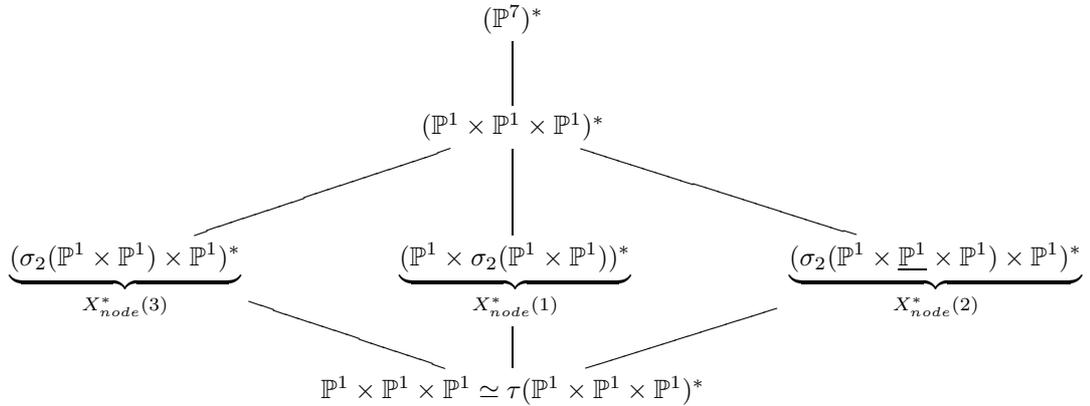

\noindent Theorem \ref{dualorbit} allows us to recover Miyake's classification of $2\times 2\times(n+1)$ quantum systems for $n\geq 1$. 
For instance a direct consequence of Table \ref{table222d} and Figure \ref{222onion} is Figure \ref{222donion} which is the geometric description
developped in \cite{My}.

\noindent Moreover with Proposition \ref{dual-Jsec} the node components are identified. 
Those node components turn out to be self-dual varieties and are geometrically described.
The same comment is true for the $2\times 2\times 3$ system. For $2\times 2\times(n+1)$ with $n\geq 4$ the interpretation of the node components is less obvious
as the dual variety of $\PP^1\times\PP^1\times\PP^n$ is not a hypersurface.

\section{Algorithms}\label{algo}
In this section, we describe an algorithmic method to identify the orbit of a given state $|\Psi\rangle$.
Consider $G$  a semi-simple Lie group and $V$ a representation of $G$. 
We know from Kac's classification \cite{Kac} which pairs $(G,V)$ have a finite number of orbits. 
A method to classify those orbits
is proposed by Vinberg \cite{Vin1,Vin2} and leads to the determination of normal forms.
Let $V_i=\CC^{n_i}$, with $1\leq i\leq k$, be $k$ complex vector spaces.  The classification of
the orbits is well known when $G$ is a product of linear groups,  $G=GL(V_1)\times\dots\times GL(V_k)$, 
acting on the tensor  space $\mathcal H:=V_1\otimes\dots\otimes V_k$ with finitely many orbits 
(or equivalently, $SL(V_1)\times\dots\times SL(V_k)$ acting on the projective  
space $\mathbb P(\mathcal{H})$).
This is the case when $k=2$ (it reduces to the classification of matrices according to their ranks, see example \ref{bipartite}) 
and $k=3$ with $n_1\times n_2\times n_3=2\times 2\times n$ and 
$n_1\times n_2\times n_3=2\times3\times n$. The normal forms are given in \cite{Pav} and in the previous sections we took
advantage of the knowledge of those normal forms to describe the orbit closures.
When $G$ acts with an infinite number of orbits, some special cases where the normal forms
depend on  parameters are solved, see for instance the $3\times 3\times 3$ case  in \cite{Nur}.\\
%
That same problem of classification of orbits under the product of linear groups 
has been  studied from the  point of view of classical invariant theory up to about 1940 (see {\emph e.g.} \cite{Hitch1,Hitch2,Hitch3,Schwar,ThrCh,Thr}) and suscited quite recently a regain of interest, motivated by potential application to quantum computing (see {\emph e.g.} \cite{Bry,BryBry,Klya,My,My2,My3}).
The classical invariant theory approach has the advantage 
to produce polynomials (invariants, covariants, concomitants) which can be used to separate the orbits.
For instance the case $2\times 2\times 2$  goes back to 1881 with the work of Le Paige \cite{LePai} where a complete 
list of covariants is described. That list of covariants allows us to distinguish the orbits. 
The $2\times 2\times 2$ case was also treated into a more precise form by Elise Schwartz \cite{Schwar}
 and independently by Saddler \cite{Sadd}. 
The result is reproduced in Sokolov's book \cite{Soko} where the state of the knowledge up to 1960 is summarized.
 The classification for the $3\times 3\times 3$ case, following the same techniques, can be found in \cite{ThrCh}. Very little is known about other cases in terms of invariants, covariants and concomittants.\\
For our purpose a state 
\[
|\Psi\rangle:=\sum_{1\leq i\leq k}\sum_{0\leq j_i\leq n_i-1}A_{j_1,\dots,j_k}|j_1\dots j_k\rangle
\]
will be assimilated to the hypermatrix $\mathcal A=\left(A_{j_1,\dots,j_k}\right)_{0\leq j_i\leq n_i-1,\forall i=1\dots k}$ 
and the first covariant of  the hypermatrix will be the multilinear form (also called the ground form):
\[
A(x,y,\dots,z):=\sum_{1\leq i\leq k}\sum_{0\leq j_i\leq n_i-1}A_{j_1,\dots,j_k}x_{j_1}y_{i_2}\dots z_{j_k}
\]
In principle the classification of the orbits can be obtained from the knowledge of the invariants, covariants and other concomitants (in the sense of classical invariant theory) of hypermatrices. This is a non-trivial task, and the result would
allow in particular to write down explicit equations for the orbits closures, i.e. for the corresponding algebraic varieties.\\
We will mainly make use of the method of Schur functions, introduced in
invariant theory by D.E. Littlewood \cite{Little}. Our strategy will be to find 
rigorously, or by a guess from numerical data, 
a generating function for the
number of covariants of any given type (the number of general concomitants is more difficult to compute), and then, guided by the series,  to try to construct
explicitly first the covariant polynomials  and hence some other concomitants by all possible methods.\\
Once the concomitants are obtained, the classification of the orbit closures 
 can be recovered by testing the nullity of the concomitants. 
The description of the algebra of concomitants is a very tedious process which needs 
(even for simple cases; for general case this is unrealizable) several hours of 
computations on a computer algebra system. Nevertheless, once the polynomials are obtained and  written 
in an appropriate way, the test is very efficient since this amounts to evaluate the polynomials
on representatives of the orbits.
\subsection{General method}
When it is possible, we first determine the Hilbert series. The Hilbert series is easier to obtain than the 
  description of the algebra and it will allow us to guide the calculations.\\
The sets of all invariants and covariants of hypermatrices of a given size are algebras $\mathrm{Inv}=S(\mathcal H)^G$ and $\mathrm{Cov}=[S(\mathcal H)\otimes S(V_1^*\oplus \dots\oplus V_k^*)]^G$ which can be graded according to the degree $d$. 
The action of $G$ on the space $\mathrm Cov$ provides an additional information: the weight of a covariant
regarded as  a relative invariant. More precisely, the weight of an invariant $F\in S^d(\mathcal H)$ is the vector $\ell=(\ell_1,\dots,\ell_k)$ such that for any $g=(g_1,\dots,g_k)\in G$
\[
g.F=(\det g_1)^{\ell_1}\dots (\det g_k)^{\ell_k}F,
\]
where $g.F$ means the image of $F$ under the natural representation of $G$ on $S^d(\mathcal H)$. 
Similarly a covariant of degree $d=(d_0,d_1,\dots,d_k)$ is a relative invariant of 
$S(\mathcal H)\otimes S(V_1^*\oplus \dots\oplus V_k^*)$. So the algebra of the  invariants and covariants can be graded 
according to both the degree $d$ and the weight $\ell$.\\
For simplicity we will consider only the space $\mathrm{Inv}(d_0)$ (resp. $\mathrm{Cov}(d)$)  of the (resp. multi) homogeneous polynomials of fixed degree $d_0$ (resp. multi-degree $d$)  and we study the multivariate Hilbert series
\[
H_{Cov}(t;u):=\sum \dim \mathrm{Cov}(d)t^{d_0}u_1^{d_1}\dots u_k^{d_k}.
\]
We show (see appendix \ref{AppHilb} for details) the Hilbert series can be written using a Cauchy function:
\[
\Pi_t[S]=\prod_{m}\left(\frac1{1-mt}\right)^{\alpha_m}
\]
where $S=\sum_{m}\alpha_mm$ is a (potentially infinite) linear combination of certain elements $m$.
For instance :
\[
\Pi_t[-u^{-1}+2v+\frac32uv]=\frac {1-tu^{-1}}{(1-tv)^2(1-tuv)^{\frac32}}. 
\]
We have:
\begin{equation}\label{HilbForm}
H_{Cov}(t;u)={\rm CT}_{v_1}\Omega^{u_1}_{\geq}\dots{\rm CT}_{v_n}\Omega^{u_n}_{\geq}B_{i_1}(u_1,v_1)\dots B_{i_n}(u_n,v_n)\Pi_t\left[A_{i_1}(u_1,v_1)\cdots A_{i_n}(u_n,v_n)\right]
\end{equation}
where  ${\rm CT}_vf(v)$ denotes the constant terms of the Laurent series $f(v)$, $\Omega_v$ is the Macmahon operator \cite{MacMa} which sends the negative power of $v$ to $0$, $A_2(u,v)=u+\frac1{u}$, $B_2(u,v)=1-{1\over u^2}$ for binary variables and $$A_2(u,v)=u+v+\frac 1{vu},\, B_3(u,v)=\left(1-{1\over vu^2}\right)\left(1-{1\over v^2u}\right)
\left(1-{v\over u}\right)$$ for ternary variables.\\
Note using equation (\ref{HilbForm}) to compute a closed form for the Hilbert series is not straightforward. The main strategy consists in decomposing the Laurent series into simple fractions, sending the fractions which contributes  negative powers of each $u$ to $0$. This is a very tedious calculation which can be performed only for the simplest cases. \\
Now, our method of construction consists in generating concomitants and testing if the dimensions of the 
graded spaces spanned by these coincide with the dimensions predicted by the Hilbert series.\\
Once constructed, we use the covariants to identify the orbit of a given form. Indeed, if $P$ is a concomitant 
then the assertion $P=0$ is invariant on an orbit. So the goal is to 
compute sufficiently many concomitants in order to distinguish the orbits. Since in our case the classification is already known, we do not need to describe completely the algebra.\\
\subsection{The case $2\times 2\times 2$}
To illustrate the method, we first apply it to the (well known) simplest non-trivial case: $k=3$ and $V_i=\mathbb C^2$. The generating series of the algebra of covariants is known :
\[
{1-t^6u_1^2u_2^2u_3^2\over (1-tu_1u_2u_3)(1-t^2u_1^2)(1-t^2u_2^2)(1-t^2u_3^2)(1-t^3u_1u_2u_3)(1-t^4)}.
\]
This suggests that the algebra is generated by a trilinear covariant of degree $1$, 
three quadratic covariants of degree $2$, a trilinear covariant of degree $3$ and a degree $4$ invariant. 
Note also that the numerator suggests a triquadratic syzygy in degree $6$.
The complete system of covariant polynomials was found by Le Paige in \cite{LePai}.  The simplest covariants is the ground form $A$. The three quadratic forms are 
\[B_x(x)=\det\left(\partial^2A\over\partial y_i\partial z_j\right)_{0\leq i,j\leq 1}, \]
\[B_y(y)=\det\left(\partial^2A\over\partial x_i\partial z_j\right)_{0\leq i,j\leq 1} \]
and
\[B_z(z)=\det\left(\partial^2A\over\partial x_i\partial y_j\right)_{0\leq i,j\leq 1}. \]
To obtain the trilinear form, one computes anyone of the three Jacobians of $A$ with one of the quadratic forms, which turn out to be the same
\[
C(x,y,z)=\left|\begin{array}{cc}\partial A\over\partial x_0& \partial A\over\partial x_1\\
\partial B_x\over\partial x_0& \partial B_x\over\partial x_1
\end{array}\right|.
\]
The three quadratic forms $B_x$, $B_y$ and $B_z$ have the same discriminant $\Delta$ which is also the hyperdeterminant \cite{GKZ} of the form. Furthermore the syzygy is
\[
C^2+\frac12B_xB_yB_z+\Delta A^2=0.
\]
With each form $A$, we associate the vector $v_A:=\langle [B_x],[B_y],[B_z],[C],[\Delta]\rangle$ where $[P]=0$ 
is $P=0$ and $[P]=1$ if $P\neq 0$. The evaluation of $v_A$ allows us to distinguish the different orbits (see Table \ref{TabOrb}).
\begin{table}[h]
\[
\begin{array}{|c|c|c|}
\hline
\rm Orbits&\rm Representatives& v_A\\\hline
\overline{\mathcal O}_{VI}&|000\rangle+|111\rangle&\langle 1,1,1,1,1\rangle
\\
\overline{\mathcal O}_{V}&|001\rangle+|010\rangle+|100\rangle&\langle 1,1,1,1,0\rangle
\\
\overline{\mathcal O}_{IV}&|111\rangle+|001\rangle& \langle 0,0,1,0,0\rangle\\
\overline{\mathcal O}_{III}&|111\rangle+|100\rangle& \langle 1,0,0,0,0\rangle\\
\overline{\mathcal O}_{II}&|111\rangle+|010\rangle& \langle 0,1,0,0,0\rangle\\
\overline{\mathcal O}_{I}&|111\rangle& \langle 0,0,0,0,0\rangle\\\hline
\end{array}
\]
\caption{\label{TabOrb} The case $2\times 2\times 2$: evaluation of $v_A$ on the orbits.}
\end{table}

\begin{rem}\rm\label{rem222} Let us compare  the orbits described in Table \ref{table222} and the covariants of Table \ref{TabOrb}. 
 Recall that in the $2\times 2\times 2$ case, the variety of separable states, corresponding to $\overline{\mathcal{O}}_I$ is $X=\PP^1\times\PP^1\times \PP^1$.
The invariant $\Delta$ is the equation of the dual variety $X^*$, the so-called Cayley hyperdeterminant but it is also
the defining equation of the tangential variety $\tau(X)$ according to Table \ref{table222}. Its singular locus corresponds
to hypermatrices $|\Psi\rangle$ such that $C(|\Psi\rangle)=0$.\\
The syzygy restricted to hypermatrices $|\Psi\rangle$ which belongs to $\tau(X)$ becomes $C^2=-\dfrac{1}{2}B_xB_yB_z$.
It tells us that the locus defined by $C=0$ is not irreducible but will be made of three components corresponding to 
the vanishing of one of the covariants $B_x$, $B_y$, $B_z$. 
To get a better understanding of the covariant $B_x$ let us consider, for a given state $|\Psi\rangle$, the projective map 
$\psi_x:\PP^1\to\PP^3$ defined by 
 $\widehat{\psi}_x:\CC^2\to\CC^2\otimes \CC^2$ with\[\widehat{\psi}_x(v)=\begin{pmatrix}
                                                                            A_{000}x_0+A_{100}x_1 & A_{001}x_0+A_{101}x_1\\
                                                                             A_{010}x_0+A_{110}x_1 & A_{011}x_0+A_{111}x_1
                                                                           \end{pmatrix} \text{ for } v=\begin{pmatrix}
x_0\\x_1 \end{pmatrix}\]
Let $\Sigma=\PP^1\times \PP^1\subset \PP^3$ be the hypersurface defined by $\det=0$ 
(the projectivization of the set of matrices of rank one in $\PP(\CC^2\otimes\CC^2)$). The definition of the covariant $B_x$ implies that $B_x(|\Psi\rangle)=0$ if and only 
if $\psi_x(\PP^1)$ is contained in $\Sigma$.
The image $\psi_x(\PP^1)$ is either a point or a line. If it is a point then $\widehat{\psi}_x(\CC^2)=f\otimes g$ and 
$|\Psi\rangle=[e\otimes f\otimes g]\in \PP^1\times \PP^1\times\PP^1$.
If $\psi_x(\PP^1)$ is a line then either $\psi_x(\PP^1)=\PP^1=\PP(\CC^2\otimes g)$ or $\psi_x(\PP^1)= \PP^1=\PP(f\otimes\CC^2)$ 
(the variety $\Sigma$ is ruled by two families of lines).
The first solution implies $|\Psi\rangle \in \sigma_2(\PP^1\times\PP^1)\times \PP^1$, i.e. $|\Psi\rangle \in \overline{\mathcal{O}}_{IV}$, the second solution
gives $|\Psi\rangle \in \sigma_2(\PP^1\times\underline{\PP^1}\times\PP^1)\times\PP^1$, i.e $|\Psi\rangle \in \overline{\mathcal{O}}_{II}$.
Therefore
 we recover geometrically that
 $B_x(|\Psi\rangle)=0 \Leftrightarrow |\Psi \rangle \in (\sigma_2(\PP^1\times \underline{\PP^1}\times\PP^1)\times\PP^1)\cup (\sigma_2(\PP^1\times \PP^1)\times\PP^1)$. 
Similarly it can be shown geometrically that $B_y$ will vanish if and only if  $|\Psi\rangle \in (\PP^1\times \sigma_2(\PP^1\times\PP^1))\cup (\sigma_2(\PP^1\times \PP^1)\times\PP^1)$
and $B_z$ will vanish if and only if $|\Psi\rangle \in(\sigma_2(\PP^1\times \underline{\PP^1}\times\PP^1)\times\PP^1)\cup (\PP^1\times\sigma_2(\PP^1\times \PP^1))$.
\end{rem}
\begin{rem}\rm
We  also give an interpretation in terms of pencils of circles in appendix \ref{circles}.
\end{rem}
\subsection{The case $2\times2\times 3$}

From (\ref{HilbForm}) we find that the Hilbert series of the algebra of covariants is
\[\begin{array}{rr}
\Omega_{\geq}^{u_1}\Omega_{\geq}^{u_2}{\rm CT}_{v_3}\Omega_{\geq}^{u_3}&\left(1-\frac1{u_1 ^2}\right)
\left(1-\frac1{u_2 ^2}\right)\left(1-\frac1{v_3u_3 ^2}\right)
\left(1-\frac1{v_3^2u_3 }\right)\left(1-\frac{v_3}{u_3}\right)\times\\
&\times\Pi_t\left[\left(u_1+\frac1{u_1}\right) 
\left(u_2+\frac1{u_2}\right)\left(u_3+v_3+\frac 1{v_3u_3}\right)\right].
\end{array}\]
The extraction of the positive part is a very tedious process which can be performed by expanding in simple fractions and erasing the fractions whose expansion as series are not Taylor.
The Hilbert series of the algebra of the covariants can be explicitly computed :
\begin{equation}\label{Hilb223}
1-t^{8}u_1^2u_2^2u_3^2\over (1-tu_1u_2u_3)(1-t^2u_3^2)(1-t^3u_1u_2)(1-t^4u_1u_2^2)(1-t^4u_1u_3^2)(1-t^6).
\end{equation}
This suggests that there are $6$ generators of degree $1$, $2$, $3$, $4$, $4$ and $6$ with only one invariant in degree $6$ and a syzygy in degree $8$.\\
The covariant of degree $1$ is just the ground form 
\[
A=\sum_{i=0}^1\sum_{j,k=0}^2a_{i,j,k}x_iy_jz_k.
\]
The covariant of degree $2$ is obtained from $A$ by elimination of variables $x$ and $y$:
\[
B:=\det\left(\partial^2A\over\partial x_i\partial y_i\right)_{0\leq i,j\leq 1}.
\]
The covariant of degree $3$ is the bilinear form
\[
C:=\left|
\begin{array}{cccc}
a_{000}&a_{100}&a_{010}&a_{110}\\
a_{001}&a_{101}&a_{011}&a_{111}\\
a_{002}&a_{102}&a_{012}&a_{112}\\
x_1y_1&-x_0y_1&-x_1y_0&x_0y_0
\end{array}
\right|=\sum_{i,j=0}^1\left(\sum_{\sigma\in\S_3}a_{i,j,\sigma(1)-1}\left(a_{0,0,\sigma(2)-1,}a_{1,1,\sigma(3)-1}-a_{1,0,\sigma(2)-1}a_{0,1,\sigma(3)-1}\right)\right)x_iy_j.
\]
The unique invariant generator is the determinant of $C$ seen as a $2\times 2$ matrix:
\[
\Delta:=\det\left(\partial^2C\over\partial x_i\partial y_j\right)
\]
To describe the two covariants in degree $4$, we recall the definition of the transvection of two multi-binary forms on the binary variables $x^{(1)}=(x^{(1)}_0,x^{(1)}_1), \dots, x^{(p)}=(x^{(p)}_0,x^{(p)}_1))$:
\[
(f,g)_{i_1,\dots,i_p}={\mathrm tr} \Omega^{i_1}_{x^{(1)}}\dots \Omega_{x^{(p)}}^{i_p}f(x'^{(1)},\dots,x'^{(p)})g(x''^{(1)},\dots,x''^{(p)}),
\]
where  $\Omega$ is the Cayley operator
\[
\Omega_x=\left|\begin{array}{cc}\partial\over \partial x'_0& \partial\over \partial x''_0
\\ \partial x'_1& \partial\over \partial x''_1\end{array}\right|
\]
and $\rm tr$ sends each variables $x', x''$ on $x$ (erases $'$ and $''$).\\
The covariant $A$ and $C$ are two bilinear forms on the binary variables $(x_0,x_1)$ and $(y_0,y_1)$. So we can apply the transvection operators and obtains two covariants in degree $4$:
\[
D_x=(A,C)_{01}\mbox{ and }D_y=(A,C)_{10}.
\]
The  evaluation of the vector $v_A:=\langle[B],[C],[D_x],[D_y],[\Delta]\rangle$ on the different orbits is reproduced 
in Table \ref{TabCov223_1}.
\begin{table}[h]
\[
\begin{array}{|c|c|}
\hline Orbit& v_A\\\hline
\overline{\mathcal O}_{VIII}&\langle1,1,1,1,1\rangle\\
\overline{\mathcal O}_{VII}&\langle1,1,1,1,0\rangle\\
\overline{\mathcal O}_{VI}&\langle1,0,0,0,0\rangle\\
\overline{\mathcal O}_{V}&\langle1,0,0,0,0\rangle\\
\overline{\mathcal O}_{IV}&\langle0,0,0,0,0\rangle\\
\overline{\mathcal O}_{III}&\langle0,0,0,0,0\rangle\\
\overline{\mathcal O}_{II}&\langle1,0,0,0,0\rangle\\
\overline{\mathcal O}_{I}&\langle0,0,0,0,0\rangle\\\hline
\end{array}
\]
\caption{\label{TabCov223_1} The case $2\times 2\times 3$: Evaluation of $v_A$ on the orbits.}
\end{table}
Note that the covariants $D_x$ and $D_y$ have no role and that $v_A$ has the same evaluation for each orbit in $\{\overline{\mathcal O}_{VI}, \overline{\mathcal O}_{V}, \overline{\mathcal O}_{II}\}$ and for each orbit in $\{\overline{\mathcal O}_{IV}, \overline{\mathcal O}_{III}, \overline{\mathcal O}_{I}\}$. So the knowledge of the covariant polynomials does 
not allow us to decide to which orbit a given state belongs. \\ \\

\noindent We need to compute more concomitant polynomials. For ternary variables, one has to consider a ternary contravariant variable $\zeta=(\zeta_0,\zeta_1,\zeta_2)$ and use an adapted version of the transvection:
\[(f,g,h)_{i,j,k}^\ell={\mathrm tr} \Omega_x^i\Omega_y^j\Omega_z^k\Omega_\zeta^\ell f(x',y',z',\xi')g(x'',y'',z'',\zeta'')h(x''',y''',z''',\zeta''')\]
where \[\Omega_p=\left|\begin{array}{cc}\partial\over \partial p'_0& \partial\over \partial p''_0
\\ \partial\over \partial p'_1& \partial\over \partial p''_1\end{array}\right|\]
for $p=x$, or $p=y$, 
 \[\Omega_p=\left|\begin{array}{ccc}\partial\over \partial p'_0& \partial\over \partial p''_0& \partial\over \partial p'''_0\\
\partial\over \partial p'_1& \partial\over \partial p''_1& \partial\over \partial p'''_1\\
\partial\over \partial p'_2& \partial\over \partial p''_2& \partial\over \partial p'''_2\end{array}\right|\]
if $p=z$ or $p=\zeta$ and $\rm tr$ is the mapping which erases the symbol $'$, $''$ and $'''$, as previously.\\
We define three concomitants:
\[
B_{x\zeta}:=(A,A,P_\zeta)^0_{1,0,1},\, B_{y\zeta}:=(A,A,P_\zeta)^0_{0,1,1}\mbox{ and }D_\zeta:=(B,B,P_\zeta)^{0}_{2,0,0},
\]
where 
\[P_\zeta:=\sum_{i=0}^2z_i\zeta_i.\]
Let $w_A=\langle[B],[B_{x\zeta}],[B_{y\zeta}],[C],[\Delta],[D_\zeta]\rangle$, we resume the evaluation of $w_A$ on the various orbits in Table \ref{TabCov223_2}.
\begin{table}[h]
\[
\begin{array}{|c|c|}
\hline Orbits& w_A\\\hline
\overline{\mathcal O}_{VIII}&\langle1,1,1,1,1,1\rangle\\
\overline{\mathcal O}_{VII}&\langle1,1,1,1,0,1\rangle\\
\overline{\mathcal O}_{VI}&\langle1,1,1,0,0,1\rangle\\
\overline{\mathcal O}_{V}&\langle1,1,1,0,0,0\rangle\\
\overline{\mathcal O}_{IV}&\langle0,1,0,0,0,0\rangle\\
\overline{\mathcal O}_{III}&\langle0,0,1,0,0,0\rangle\\
\overline{\mathcal O}_{II}&\langle1,0,0,0,0,0\rangle\\
\overline{\mathcal O}_{I}&\langle0,0,0,0,0,0\rangle\\\hline
\end{array}
\]
\caption{\label{TabCov223_2} The case $2\times 2\times 3$: evaluation of $w_A$ on the orbits.}
\end{table}
\begin{rem}\rm
 In the $2\times 2\times 3$ case, the orbit $\mathcal{O}_I$ is the Segre product $X=\PP^1\times\PP^1\times \PP^2$. The invariant $\Delta$ is
the hyperdeterminant in the sense of \cite{GKZ} of format $2\times 2\times 3$, i.e. the equation of the dual variety $X^*$.
According to Table \ref{table223}, the invariant $\Delta$ can also be interpreted as the equation of $J(X,\overline{\mathcal{O}}_{IV})$.
Like in the $2\times 2\times 2$ case, the covariant $C$ vanishes on the suborbits of the hypersuface defined by $\Delta=0$, i.e. $C(|\Psi\rangle)=0$ means $|\Psi\rangle$ is 
 a singular point of $\Delta=0$. 
The covariant $C$ admits also the following interpretation in term of secant varieties:
\[C(|\Psi\rangle)=0\Leftrightarrow |\Psi\rangle \in \sigma_2(\PP^1\times\PP^1\times \PP^2)\]
To prove this assertion, let us consider $|\Psi\rangle$ as a projective map $\psi:\PP^2\to \PP^3$ defined by the linear 
map $\widehat{\psi}:\CC^3\to \CC^2\otimes \CC^2$ given by 
$\widehat{\psi}(v)=M_0\otimes e_0 ^*(v)+M_1\otimes e_1^*(v)+M_2\otimes e_2^*(v)$, with 
$M_k=(A_{ijk})_{0\leq i,j\leq 1}\in \CC^2\otimes \CC^2$ for $k=0,1,2$ and $e_i^*$ the dual basis of $\CC^2$ (see remark \ref{rem222}). 
The linear map $ \widehat{\psi}$ is of rank $3$ precisely when $C\neq 0$. But 
the rank of $\widehat{|\Psi\rangle}\in \CC^2\otimes\CC^2\otimes \CC^3$ and the rank of $\widehat\psi$ satisfy by
 construction \[\text{rank}(\widehat{|\Psi\rangle})\geq \text{rank}(\widehat\psi)\]
If $|\Psi\rangle$ is a limit of (the projectivization of) tensors of rank less than $2$, by continuity of $C$ we have $C(|\Psi\rangle)=0$, i.e. 
$|\Psi\rangle \in \sigma_2(\PP^1\times\PP^1\times\PP^2)$ implies $C(|\Psi\rangle)=0$.\\
On the other hand if $C(|\Psi\rangle)=0$ we can assume without loss of generality that $\widehat\psi=M_0\otimes e_0^*+M_1\otimes e_1^*$.
If $M_0=\lambda M_1$ we can write the map $\widehat{\psi}$ as $\widehat\psi=M_0\otimes e_0^*$, i.e. $|\Psi\rangle=(e_0\otimes e_0+e_1\otimes e_1)\otimes e_0$ and 
$|\Psi\rangle$ is a point of $\sigma_2(\PP^1\times\PP^1)\times\PP^2$. 
If $M_0$ and $M_1$ are not colinear, these two matrices define a line after projectivization, i.e.  
$\PP^1=\PP(\lambda M_0+\mu M_1)\subset \PP^3=\PP(\CC^2\otimes\CC^2)$. 
Again consider $\Sigma=\PP^1\times\PP^1$ the hypersurface of $\PP(\CC^2\otimes\CC^2)$ defined by $\det=0$.
If $\det(\lambda M_0+\mu M_1)=0$, i.e. the projectivized line defined by $M_0$ and $M_1$ is contained is $\Sigma$, then we can assume either
$M_0=e_0\otimes e_0$ and $M_1=e_0\otimes e_1$ or $M_0=e_0\otimes e_0$ and $M_1=e_1\otimes e_0$ (both matrices are of rank $1$ and their linear combination is of rank $1$) 
and therefore either $\widehat{|\Psi\rangle}=e_0\otimes e_0\otimes e_0+e_0\otimes e_1\otimes e_1$
 and $|\Psi\rangle$ belongs to $\PP^1\times\sigma_2(\PP^1\times\PP^2)$ or 
$\widehat{|\Psi\rangle}=e_0\otimes e_0\otimes e_0+e_1\otimes e_0\otimes e_1$ and $[\Psi\rangle$  belongs to $\sigma_2(\PP^1\times\underline{\PP^1}\times\PP^2)\times \PP^1$ 
(like in example \ref{rem222} the two possibilities
correspond to the two families of lines of $\Sigma=\PP^1\times\PP^1$).\\
If $\det(\lambda M_1+\mu M_2)\neq 0$ for some values $\lambda,\mu$, i.e. the line intersects $\Sigma$, 
then we can assume $M_0$ is a $2\times 2$ matrix of rank $2$ and $M_1$ a matrix of rank $1$. 
There are two cases to consider:
\begin{itemize}
 \item  The line is tangent to $\Sigma$ and we can assume $M_0=e_0\otimes e_1+e_1\otimes e_0$ and 
$M_1=e_0\otimes e_0$. Then $\widehat{|\Psi\rangle}=(e_0\otimes e_1+e_1\otimes e_0)\otimes e_0+e_0\otimes e_0\otimes e_1=e_0\otimes e_1\otimes e_0+e_1\otimes e_0\otimes e_0+e_0\otimes e_0\otimes e_1=\frac{1}{\varepsilon}(-e_0\otimes e_0\otimes e_0+(e_0+\varepsilon e_1)\otimes (e_0+\varepsilon e_1)\otimes (e_0+\varepsilon e_1))$.
From $\widehat{|\Psi\rangle_\epsilon}=\frac{1}{\varepsilon}(-e_0\otimes e_0\otimes e_0+(e_0+\varepsilon e_1)\otimes (e_0+\varepsilon e_1)\otimes (e_0+\varepsilon e_1))$ we see that $\widehat{|\Psi\rangle}$ is a limit  of tensor of rank $2$ when $\varepsilon\to 0$, i.e. $|\Psi\rangle$ is in the closure of $\sigma_2(\PP^1\times\PP^1\times\PP^2)$.
\item The line is secant to $\Sigma$ and we can assume $M_0=e_0\otimes e_0+e_1\otimes e_1$ and $M_2=e_0\otimes e_0$. Then $\widehat{|\Psi\rangle}=e_0\otimes e_0\otimes e_0+e_1\otimes e_1\otimes e_1$, i.e. $|\Psi\rangle \in \sigma_2(\PP^1\times\PP^1\times\PP^2)$.
\end{itemize}

\noindent The covariant $B$ is the analogue of the covariants $B_x, B_y, B_z$ in the $2\times 2\times 2$ case, 
i.e. $B(|\Psi\rangle)=\det(\widehat{\psi}(\CC^2))$. The vanishing 
of $B(|\Psi\rangle)$ implies  $\psi(\PP^2)$ belongs to $\Sigma$.
In particular one sees that $B$ vanishes  only if $C=0$ ($C\neq 0$ implies $\psi(\PP^2)$ is a plane and therefore can not be contained in $\Sigma$).
Like in the $2\times 2\times 2$ case there will be three cases corresponding to $B=0$ which are:
\begin{itemize}
 \item $\psi(\PP^2)$ is a point of 
$\Sigma$ (orbit $\overline{\mathcal{O}}_I=\PP^1\times\PP^1\times\PP^2$), 
 \item $\psi(\PP^2)$ is a line of $\Sigma$, i.e. 
\begin{itemize}
\item either
$\psi(\PP^2)=\PP(f\otimes \CC^2)$ (orbit $\overline{\mathcal{O}}_{IV}=\PP^1\times \sigma_2(\PP^1\times\PP^2)$) 
\item or $\psi(\PP^2)=\PP(\CC^2\otimes g)$ 
(orbit $\overline{\mathcal{O}}_{III}=\sigma_2(\PP^1\times \underline{\PP^1}\times\PP^2)\times\PP^1$). 
\end{itemize}
\end{itemize}
Those different cases are distinguished by the concomitants $B_{x\zeta}$ and $B_{y\zeta}$.
If $C=0$ and $B\neq 0$, then there are three different cases
\begin{itemize}
 \item $\psi(\PP^2)$ is a point and $\psi(\PP^2)\notin \Sigma$ (orbit $\overline{\mathcal{O}}_{II}$)
 \item $\psi(\PP^2)$ is a line tangent to $\Sigma$ (orbit $\overline{\mathcal{O}}_{V}$)
 \item $\psi(\PP^2)$ is a line secant to $\Sigma$ (orbit $\overline{\mathcal{O}}_{VI}$)
\end{itemize}
Finaly when $C\neq 0$ then $\psi(\PP^2)$ is a plane which could be tangent to $\Sigma$, $\Delta=0$  
(orbit $\overline{\mathcal{O}}_{VII}$) or secant, $\Delta\neq 0$ (orbit $\overline{\mathcal{O}}_{VIII}$).
\end{rem}

\begin{rem}\rm
In appendix  \ref{circles} we give an interpretation in terms of linear complex of circles.
\end{rem}

\subsection{The case $2\times 3\times 3$}
Again we start by computing the multivariate Hilbert series:
\[\begin{array}{rr}\displaystyle
\Omega_{\geq}^{u_1}{\rm CT}_{v_2}\Omega_{\geq}^{u_2}{\rm CT}_{v_3}\Omega_{\geq}^{u_3}&\displaystyle\left(1-\frac1{u_1 ^2}\right)
\left(1-\frac1{v_2u_2 ^2}\right)\left(1-\frac1{v_3u_3 ^2}\right)
\left(1-\frac1{u_2v_2 ^2}\right)\left(1-\frac1{u_3v_3 ^2}\right)\left(1-\frac{v_2}{u_2}\right)
\times\\&\displaystyle\times\left((1-\frac{v_3}{u_3}\right)\Pi_t\left[\left(u_1+\frac1{u_1}\right) 
\left(u_2+{v_2}+\frac1{v_2u_2}\right)\left(u_3+v_3+\frac1{v_3u_3}\right)\right].
\end{array}\]
 We find
\begin{equation}
P(t;u_1,u_2,u_3)\over (1-tu_1u_2u_3)(1-u_1^3t^3)(1-u_2u_3t^4)(1-u_2^3t^6)
(1-u_3^3t^6)(1-u_2^3u_2^3t^6)(1-u_1^2t^6)((1-t^{12})
\end{equation}
with \[\begin{array}{rcl}
P(t;u_1,u_2,u_3)&=&-{t}^{26}{u_{{1}}}^{4}{u_{{2}}}^{5}{u_{{3}}}^{5}-{t}^{22}{u_{{1}}}^{2}
{u_{{2}}}^{4}{u_{{3}}}^{4}-{t}^{21}{u_{{1}}}^{3}{u_{{2}}}^{3}{u_{{3}}}
^{3}-{t}^{19}{u_{{1}}}^{3}{u_{{2}}}^{4}{u_{{3}}}^{4}-{t}^{18}{u_{{1}}}
^{4}{u_{{2}}}^{3}{u_{{3}}}^{3}\\
&&-{t}^{17}{u_{{1}}}{u_{{2}}}^{5}u_{{3
}}^{5}- \left( u_{{1}}-1 \right)  \left( u_{{1}}+1 \right) {t}^{16}{u_{{3}
}}^{4}{u_{{2}}}^{4}- \left( u_{{1}}-1 \right)  \left( u_{{1}}+1
 \right) {t}^{15}{u_{{3}}}^{3}{u_{{2}}}^{3}u_{{1}}\\&&- \left( u_{{1}}-1
 \right)  \left( u_{{1}}+1 \right) {t}^{11}{u_{{3}}}^{2}{u_{{2}}}^{2}u
_{{1}}
- \left( u_{{1}}-1 \right)  \left( u_{{1}}+1 \right) {t}^{10}u_{
{3}}u_{{2}}{u_{{1}}}^{2}\\&&+{t}^{9}{u_{{1}}}^{3}+{t}^{8}{u_{{3}}}^{2}{u_{
{2}}}^{2}+{t}^{7}u_{{3}}u_{{2}}u_{{1}}+{t}^{5}{u_{{3}}}^{2}{u_{{2}}}^{
2}u_{{1}}+{t}^{4}u_{{3}}u_{{2}}{u_{{1}}}^{2}+1.
\end{array}
\]
This suggests a very complicated description of the algebra. Nevertheless, as in the other cases, we will use  only a part of the concomitant polynomials. The degree $3$ generator is a cubic binary form obtained from the ground form 
$A:=\sum_{i=0}^1\sum_{j,k=0}^2a_{ijk}x_iy_jz_k$ by:
\[
C_x:=\det\left(\partial^2A\over\partial y_j\partial z_j\right)_{j,k=0..2}.
\]
For a cubic binary form, the algebra of covariants is well known. Its Hilbert series is
\[
{1-u_1^6t^6\over (1-tu_1^3)(1-u_1^2t^2)(1-u_1^3t^3)(1-t^4)}.
\]
Considering a generic cubic binary form
\[
a:=a_0x_0^3+a_1x_1x_0^2+a_2x_1^2x_0+a_3x_1^3,
\]
we compute a covariant of degree $2$, the Hessian:
\begin{equation}\label{defbcub}\begin{array}{rcl}
b&:=&\det\left(\partial^2 a\over \partial x_i\partial x_j\right)_{0\leq i,j\leq 1}
\\&=&(3a_0a_2-a_1^2)x_0^2+(9a_0a_3-a_1a_2)x_1x_0+(3a_1a_3-a_2)^2x_1^2.\end{array}
\end{equation}
The discriminant of this quadratic form is the only invariant generator:
\begin{equation}\label{defdcub}
d:=4(3a_0a_2-a_1^2)(3a_1a_3-a_2^2)-(9a_0a_3-a_1a_2)^2.
\end{equation}
We need also to compute the covariant of degree $3$:
\begin{equation}\label{defccub}\begin{array}{rcl}c&:=&  \left( 6\,z_{{0}}{\it a_0}\,{\it a_2}-2\,z_{{0}}{{\it a_1}}^{2}
-{\it a_1}\,z_{{1}}{\it a_2}+9\,{\it a_0}\,{\it a_3}\,z_{{1}}
 \right)  \left( {\it a_1}\,{z_{{0}}}^{2}+2\,{\it a_2}\,z_{{1}}z_{{0
}}+3\,{\it a_3}\,{z_{{1}}}^{2} \right) \\&&- \left( -z_{{0}}{\it a_1}
\,{\it a_2}+9\,z_{{0}}{\it a_0}\,{\it a_3}+6\,z_{{1}}{\it a_1}\,{
\it a_3}-2\,z_{{1}}{{\it a_2}}^{2} \right)  \left( 3\,{\it a_0}\,{z
_{{0}}}^{2}+2\,{\it a_1}\,z_{{1}}z_{{0}}+{\it a_2}\,{z_{{1}}}^{2}
 \right). 
\end{array} \end{equation}
Note that  we have the syzygy: $9b+512c+128da^2=0$.\\
Replacing each $a_i$ by the coefficient of $x_0^ix_1^{3-i}$ in $C_x$ in eq. (\ref{defbcub},\ref{defdcub},\ref{defccub}), that is,
\[
a_0=\left|
\begin{array}{ccc}
a_{000}&a_{001}&a_{002}\\
a_{010}&a_{011}&a_{012}\\
a_{020}&a_{021}&a_{022}
\end{array}
\right|,
\]
\[
a_1=\left|
\begin{array}{ccc}
a_{000}&a_{001}&a_{002}\\
a_{010}&a_{011}&a_{012}\\
a_{120}&a_{121}&a_{122}
\end{array}
\right|+\left|\begin{array}{ccc}
a_{000}&a_{001}&a_{002}\\
a_{110}&a_{111}&a_{112}\\
a_{020}&a_{021}&a_{022}
\end{array}
\right|+\left|\begin{array}{ccc}
a_{100}&a_{101}&a_{102}\\
a_{010}&a_{011}&a_{012}\\
a_{020}&a_{021}&a_{022}
\end{array}
\right|,
\]
\[
a_2=\left|
\begin{array}{ccc}
a_{000}&a_{001}&a_{002}\\
a_{110}&a_{111}&a_{112}\\
a_{120}&a_{121}&a_{122}
\end{array}
\right|+\left|\begin{array}{ccc}
a_{100}&a_{101}&a_{102}\\
a_{110}&a_{111}&a_{112}\\
a_{020}&a_{021}&a_{022}
\end{array}
\right|+\left|\begin{array}{ccc}
a_{100}&a_{101}&a_{102}\\
a_{010}&a_{011}&a_{012}\\
a_{120}&a_{121}&a_{122}
\end{array}
\right|
\]
and
\[
a_3=\left|
\begin{array}{ccc}
a_{100}&a_{101}&a_{102}\\
a_{110}&a_{111}&a_{112}\\
a_{120}&a_{121}&a_{122}
\end{array}
\right|;\]
 we obtain one quadratic covariant $F_x$ in degree $6$, a cubic covariant $I_x$ in degree $9$ and the invariant $\Delta$ in degree $12$.\\
We observe that $F_x=0$ if and only if $I_x=0$, so we have only to evaluate the vector $v_A=\langle[C_x],[F_x],[\Delta]\rangle$ on the representative of the different orbits.
These covariants are not sufficient to discriminate between the orbits but they allow us to partition the set of the orbits into $4$ sets (see Table \ref{233_1}).
\begin{table}[h]
\[
\begin{array}{|c|c|}
\hline Orbits& v_A\\\hline
\overline{\mathcal O}_{XVII}&\langle1,1,1\rangle\\
\overline{\mathcal O}_{XVI},\overline{\mathcal O}_{XIV}&\langle1,1,0\rangle\\
\overline{\mathcal O}_{XV},\overline{\mathcal O}_{XIII},\overline{\mathcal O}_{IX}&\langle1,0,0\rangle\\
\overline{\mathcal O}_{I},\dots,\overline{\mathcal O}_{VIII},\overline{\mathcal O}_{X},
\overline{\mathcal O}_{XI}, \overline{\mathcal O}_{XII}&\langle0,0,0\rangle\\\hline
\end{array}
\]
\caption{\label{233_1} Evaluation of $v_A$ on the orbits: the case $2\times 3\times 3$.}
\end{table}
The other covariant polynomials are more complicated to understand in an algebraic way. For instance, we will use the two following degree 4 covariants:
\[
D_{xyz}:=(C_x,A)^{100}
\]
and
\[
D_{yz}:=\mathrm{tr}\, \left(\Omega_x\mathrm{tr'}_x\Omega'_{x}\Omega_z\Omega_yA(x',y',z')^2A(x'',y'',z'')A(x''',y''',z''')\right)
\]
where $\Omega'_x:=\left|\begin{array}{cc}\partial\over \partial x''_0& \partial\over \partial x'''_0
\\ \partial x''_1& \partial\over \partial x'''_1\end{array}\right|$ and $\mathrm{tr}'_x$ sends $x''_i$ and $x'''_i$ to $x''_i$.
The covariants $D_{yz}$ and $D_{xyz}$ are both bilinear in the ternary variables $y=\{y_0,y_1,y_2\}$ and $z=\{z_0,z_1,z_2\}$ and $D_{xyz}$ is quadratic in the binary variable $x=\{x_0,x_1\}$.  
These covariants are used to separate the orbits $\overline{\mathcal O}_{XV},\overline{\mathcal O}_{XIII}$ and $\overline{\mathcal O}_{IX}$ (see Table \ref{XV-XIII-IX}).
\begin{table}[h]
\[
\begin{array}{|c|c|}
\hline Orbits& \langle [D_{yz}],[D_{xyz}]\rangle\\\hline
\overline{\mathcal O}_{XV}&\langle1,1\rangle\\
\overline{\mathcal O}_{XIII}&\langle0,1\rangle\\
\overline{\mathcal O}_{IX}&\langle0,0\rangle\\\hline
\end{array}
\]
\caption{\label{XV-XIII-IX} Evaluation of $\langle [D_{yz}],[D_{xyz}]\rangle$ on the orbits $\overline{\mathcal O}_{XV},\overline{\mathcal O}_{XIII}$ and $\overline{\mathcal O}_{IX}$.}
\end{table}
Define also
\[
F_y:=\mathrm{tr}\Omega_x\Omega_zA(x',y',z')A(x'',y'',z'')D_{yz}(y''',z''').
\]
This polynomial vanishes on  $\overline{\mathcal O}_{XIV}$ but not on $\overline{\mathcal O}_{XVI}$. It remains to separate the orbits $\overline{\mathcal O}_{I},\dots,\overline{\mathcal O}_{VIII}$, $\overline{\mathcal O}_{X}$,
$\overline{\mathcal O}_{XI}$ 	and $\overline{\mathcal O}_{XII}$.\\
We need to introduce some covariant polynomials:
\[
B_{z\eta}:=\mathrm{tr}\Omega_y\Omega_xA(x',y',z')A(x'',y'',z'')P(y''',\eta'''),
\]
and
\[
B_{y\zeta}:=\mathrm{tr}\Omega_z\Omega_xA(x',y',z')A(x'',y'',z'')P(z''',\zeta''').
\]
These polynomials allows to discriminate between the orbits  $\overline{\mathcal O}_{I}$, $\overline{\mathcal O}_{II}$ and $\overline{\mathcal O}_{III}$ (see Table \ref{I-II-III}).
\begin{table}[h]
\[
\begin{array}{|c|c|}
\hline Orbits& \langle [B_{y\zeta}],[B_{z\eta}]\rangle\\\hline
\overline{\mathcal O}_{III}&\langle0,1\rangle\\
\overline{\mathcal O}_{II}&\langle1,0\rangle\\
\overline{\mathcal O}_{I}&\langle0,0\rangle\\\hline
\end{array}
\]
\caption{\label{I-II-III} Evaluation of $\langle [B_{y\zeta}],[B_{z\eta}]\rangle$ on the orbits $\overline{\mathcal O}_{I},\overline{\mathcal O}_{II}$ and $\overline{\mathcal O}_{III}$.}
\end{table}
In degree $2$, we define also the concomitant:
\[
B_{x\eta\zeta}:=(A,A,P(y,\eta)P(z,\zeta))^{011}.
\]
In degree $4$:
\[
D_{\eta\zeta}:=(B_{x\eta\zeta},B_{x\eta\zeta})^{200},
\]
and
\[
D_{yz\eta\zeta}:=\mathrm{tr }\Omega_y\Omega_z(\Omega_xA(x',y',z')A(x'',y'',z''))^2P(y''',\eta''')P(z''',\zeta''').
\]

\noindent Also in degree $6$:
\[
F_{\eta}:=(B_{z\eta},B_{z\eta},B_{z\eta})^{002},
\]
\[
F_{\zeta}:=(B_{y\zeta},B_{y\zeta},B_{y\zeta})^{002}.
\]
Finally, in degree $8$:
\[
H_{xyz\eta}:=\mathrm{tr }\Omega_\zeta D_{yz\eta\zeta}(y',z',\eta',\zeta')
B_{z\eta}(z'',\eta'')B_{x\eta\zeta}(x''',\eta''',\zeta''')
\]
and
\[
H_{xyz\zeta}:=\mathrm{tr }\Omega_\eta D_{yz\eta\zeta}(y',z',\eta',\zeta')
B_{y\zeta}(y'',\zeta'')B_{x\eta\zeta}(x''',\eta''',\zeta''')
\]
We define $v'_A:=\langle[D_{yz\eta\zeta}],[B_{x\eta\zeta}],[D_{\eta\zeta}],[D_{yz}],[F_\eta],[F_\zeta],[H_{xyz\zeta}],[H_{xyz\eta}] \rangle$. Table \ref{IV-XII} gives the evaluation of $v'_A$ on the remaining orbits.

\begin{table}[h]
\[
\begin{array}{|c|c|}
\hline Orbits& v'_A\\\hline
\overline{\mathcal O}_{XII}&\langle1,1,1,0,1,0,0,1\rangle\\
\overline{\mathcal O}_{XI}&\langle1,1,1,1,0,0,1,1\rangle\\
\overline{\mathcal O}_{X}&\langle1,1,1,0,0,0,0,1\rangle\\
\overline{\mathcal O}_{VIII}&\langle1,1,1,0,0,1,1,0\rangle\\
\overline{\mathcal O}_{VII}&\langle1,1,1,0,0,0,1,0\rangle\\
\overline{\mathcal O}_{VI}&\langle1,1,1,0,0,0,0,0\rangle\\
\overline{\mathcal O}_{V}&\langle 1,1,0,0,0,0,0,0\rangle\\
\overline{\mathcal O}_{IV}&\langle0,1,0,0,0,0,0,0\rangle\\\hline
\end{array}
\]
\caption{\label{IV-XII} Evaluation of $v'_A$ on the orbits $\overline{\mathcal O}_{IV},\dots,\overline{\mathcal O}_{VIII}$, $\overline{\mathcal O}_{X}$,
$\overline{\mathcal O}_{XI}$ 	and $\overline{\mathcal O}_{XII}$.}
\end{table}
We summarize the results of this section in Table \ref{table233cov}, setting
\[
w_A:=\langle [D_{yz\eta\zeta}],[B_{x\eta\zeta}],[B_{y\zeta}], [B_{z\eta}],[C_x], [D_{\eta\zeta}],[D_{yz}],[F_y],[F_x],[F_\eta],[F_\zeta],[H_{xyz\zeta}],[H_{xyz\eta}],[\Delta]\rangle.
\]
\begin{table}[h]
\[
\begin{array}{|c|c|}
\hline Orbits& w_A\\\hline
\overline{\mathcal O}_{XVII}&\langle   1,1,1,1,1,1,1,1,1,1,1,1,1,1\rangle\\
\overline{\mathcal O}_{XVI}&\langle    1,1,1,1,1,1,1,1,1,1,1,1,1,0\rangle\\
\overline{\mathcal O}_{XV}&\langle     1,1,1,1,1,1,1,1,0,1,1,1,1,0\rangle\\
\overline{\mathcal O}_{XIV}&\langle    1,1,1,1,1,1,1,0,1,0,0,1,1,0\rangle\\
\overline{\mathcal O}_{XIII}&\langle   1,1,1,1,1,1,0,0,0,0,0,1,1,0\rangle\\
\overline{\mathcal O}_{XII}&\langle    1,1,1,1,0,1,0,0,0,1,0,0,1,0\rangle\\
\overline{\mathcal O}_{XI}&\langle     1,1,1,1,0,1,1,1,0,0,0,1,1,0\rangle\\
\overline{\mathcal O}_{X}&\langle      1,1,1,1,0,1,0,0,0,0,0,0,1,0\rangle\\
\overline{\mathcal O}_{IX}&\langle     0,1,0,0,1,0,0,0,0,0,0,0,0,0\rangle\\
\overline{\mathcal O}_{VIII}&\langle   1,1,1,1,0,1,0,0,0,0,1,1,0,0\rangle\\
\overline{\mathcal O}_{VII}&\langle    1,1,1,1,0,1,0,0,0,0,0,1,0,0\rangle\\
\overline{\mathcal O}_{VI}&\langle     1,1,1,1,0,1,0,0,0,0,0,0,0,0\rangle\\
\overline{\mathcal O}_{V}&\langle      1,1,1,1,0,0,0,0,0,0,0,0,0,0\rangle\\
\overline{\mathcal O}_{IV}&\langle     0,1,0,0,0,0,0,0,0,0,0,0,0,0\rangle\\
\overline{\mathcal O}_{III}&\langle    0,0,1,0,0,0,0,0,0,0,0,0,0,0\rangle\\
\overline{\mathcal O}_{II}&\langle     0,0,0,1,0,0,0,0,0,0,0,0,0,0\rangle\\
\overline{\mathcal O}_{I}&\langle      0,0,0,0,0,0,0,0,0,0,0,0,0,0\rangle\\\hline
\end{array}
\]
\caption{\label{table233cov} Case $2\times3\times3$: Evaluation of $w_A$ on the orbits.}
\end{table}

\begin{rem}\rm
The comparaison between varieties of Table \ref{table233} and \ref{table233cov} give 
the following interpretations for $\Delta$, $C_x$, $B_{y\zeta}$, $B_{z\eta}$:\\
 Again the only invariant polynomial $\Delta$ can be considered as
the equation of the dual of $X=\PP^1\times \PP^2\times \PP^2$ or as the  equation of the join 
$J(X,\tau(X))$.\\
The covariant $C_x$ is defined by $C_x(|\Psi\rangle)=\det(\widehat{\psi}_x(\CC^2))$ where
$\widehat{\psi}_x:\CC^2\to \CC^3\otimes \CC^3$, is given by  $\widehat{\psi}_x=e_0^*\otimes M_0+e_1^*\otimes M_1$, with $M_i$ 
$3\times 3$ matrices.
Thus $C_x(|\Psi\rangle)\neq 0$ means $\psi_x(\PP^1)$ is not contained in $\Sigma=\sigma(\PP^2\times\PP^2)\subset \PP^8$, the hypersurface defined by $\det=0$.
This is the case when $|\Psi\rangle$ is a general point of  $\overline{\mathcal{O}}_{IX}=\PP^1\times\PP^8$. Thus $C_x$
does not vanish on all orbits containing $\overline{\mathcal{O}}_{IX}$, i.e. 
orbits $\overline{\mathcal O}_{XIII},\overline{\mathcal O}_{XIV},\overline{\mathcal O}_{XV},\overline{\mathcal O}_{XVI},\overline{\mathcal{O}}_{XVII}$ (see Figure \ref{233onion}).
On the other hand when $C_x(|\Psi\rangle)=0$, then $\psi_x(\PP^1)$ is a 
subset of $\Sigma=\sigma_2(\PP^2\times\PP^2)\subset \PP^8$.\\
The image  $\psi_x(\PP^1)$ is either a point or a line. 
It is a point when $B_{y\zeta}(|\Psi\rangle)=0$ and $B_{z\eta}(|\Psi\rangle)=0$ 
i.e. the matrices $M_0$ and $M_1$ which define $\psi_x$ are colinear.
The discussion on $C_x$ and $B_{y\zeta}$, $B_{z\eta}$ gives the following possibilities:
\begin{itemize}
 \item if $B_{y\zeta}(|\Psi\rangle)=B_{z\eta}(|\Psi\rangle)=0$ ($\psi_x(\PP^1)$ is a point of $\PP^8$)
\begin{itemize}
 \item then $C_x(|\Psi\rangle)=0$ (i.e. $\psi(\PP^1)\in \Sigma$), gives two possibilites:
\begin{itemize}
\item $\psi_x(\PP^1)$ is a point of $\PP^2\times\PP^2$ (orbit $\overline{\mathcal{O}}_I$)
\item $\psi_x(\PP^1)$ is a point of $\Sigma$ (orbit $\overline{\mathcal{O}}_{IV}$)
\end{itemize}
\item and $C_x(|\Psi\rangle)\neq 0$ (i.e. $\psi(\PP^1)\notin \Sigma$), gives one possibility:
\begin{itemize}
\item $\psi_x(\PP^1)$ is a general point of $\PP^8\setminus\Sigma$ (orbit $\overline{\mathcal{O}}_{IX}$).
\end{itemize}
\end{itemize}
\item if $B_{y\zeta}(|\Psi\rangle)\neq 0$ or $B_{z\eta}(|\Psi\rangle)\neq 0$ ($\psi_x(\PP^1)$ is line in $\PP^8$)
\begin{itemize}
 \item then $C_x(|\Psi\rangle)=0$ (i.e. $\psi(\PP^1)\subset \Sigma$), gives 9 possibilites:
\begin{itemize}
\item $\psi_x(\PP^1)$ is a line of $\PP^2\times\PP^2$ (orbits $\overline{\mathcal{O}}_{II}$ and $\overline{\mathcal{O}}_{III}$)
\item $\psi_x(\PP^1)$ is a line tangent to $\PP^2\times\PP^2$  (orbit $\overline{\mathcal{O}}_{V}$)
\item $\psi_x(\PP^1)$ is a line secant to $\PP^2\times \PP^2$ (orbit $\overline{\mathcal{O}}_{VI}$)
\item $\psi_x(\PP^1)$ is a line of $\Sigma$ (orbits  $\overline{\mathcal{O}}_{VIII}$,  
$\overline{\mathcal{O}}_{X}$,  $\overline{\mathcal{O}}_{XI}$,  $\overline{\mathcal{O}}_{XII}$)
\end{itemize}
\item and $C_x(|\Psi\rangle)\neq 0$ (i.e. $\psi(\PP^1)\cap \Sigma=\emptyset$), gives 5 possibilities:
\begin{itemize}
\item $\psi_x(\PP^1)$ is a line of $\PP^8\setminus\Sigma$ (orbit $\overline{\mathcal{O}}_{XIII}$, $\overline{\mathcal{O}}_{XIV}$,
$\overline{\mathcal{O}}_{XV}$, $\overline{\mathcal{O}}_{XVI}$, $\overline{\mathcal{O}}_{XVII}$).
\end{itemize}
\end{itemize}
\end{itemize}
The covariants and concomitants allow us to distinguish the different positions of the lines but a priori geometric 
interpretations of those polynomials are far from being obvious.
\end{rem}


\noindent Note Table \ref{table233cov} have been computed using {\rm Maple} programs.\\
 The sources are available at {\tt http://www-igm.univ-mlv.fr/$\sim$luque/form233.txt}.

\section{Conclusion}\label{conclusion}
In this paper we proposed an alternative approch to the geometric descriptions of entanglement given by Miyake in \cite{My,My2}.
The idea was to use auxiliary varieties such as join, tangent and secant  varieties instead of a description by dual varieties. 
The introduction of the secant and tangential varieties brought a more precise description of the singular locus of the dual varieties as we were able to 
interpret the singular components of $X^*$ as dual varieties of the stratification by join and tangential varieties. 
We also detailed the geometric description of the entangled states for 
the $2\times3\times 3$ quantum system.  
Both descriptions of entanglement, by join and tangential varieties or dual varieties, are  
equivalent as long as we deal with group actions with finetly many orbits. However challenging problems in QIT
 start  with 
quantum systems with infinitely many orbits. 
For example in the  case of the $3\times3\times3$ system or the $2\times2\times2\times2$ 
system there does not seem to be a complete consensus on 
what an entangled state is mathematicaly (different papers announce different numbers of entangled states under different definitions  \cite{VDMV, BDD,LLS}).
We believe that the approach by secant and tangential varieties could bring interesting perspectives 
for the geometric description of entanglement in QIT. 
For instance the recent work of Buczy\'nsky and Landsberg on the third secant varieties of the Segre product of three 
projective spaces provides useful results to describe the points lying in the closure of $\sigma_3(\PP^{k_1}\times\PP^{k_2}\times\PP^{k_3})$.
But smooth points of $\sigma_3(\PP^{k_1}\times\PP^{k_2}\times\PP^{k_3})$ correspond to a state of type
$|000\rangle+|111\rangle+|222\rangle$, i.e. generalize the GHZ state. As noticed in \cite{hey} the results of \cite{CGG} on the geometry of 
$\sigma_s(\underbrace{\PP^1\times\dots\times\PP^1}_{\text{n times}})$ should lead to a better 
understanding of the quantification of entanglement for $n$-qubits.

\noindent  Another way to study mathematicaly entanglement is to look for invariant polynomials under the SLOCC group actions \cite{BLT,LT}.
These polynomials allow
 one to distinguish the different states as we did in Section \ref{algo} where 
we introduced an algorithmic method
to identify the orbit of a given state for the quantum systems of this paper. 
In this sense the dual varieties approach provides also invariant polynomials as the hyperdeterminant,
in the sense \cite{GKZ}, is SLOCC invariant. But the study of the defining
equations of the secant and tangential varieties is also a topic of interest for algebraic geometers \cite{CGG2,LM}.
Looking for geometric interpretations of the 
covariants we made some connection between the vanishing of some 
these polynomials and the defining equations of the secant variety. We believe that the geometric understanding of 
the polynomials produced by classical invariant theory techniques in the context of auxiliary varieties needs to be better understood.
\noindent The study of the geometry of secant and tangent varieties of Segre varieties should therefore help to understand the structure
 of multipartite entanglement both
from the orbit stratifications point of view and from the invariant polynomials perspective.

\appendix
\section{Hilbert series\label{AppHilb}}
In this section, we use extensively symmetric functions (and in particular Schur functions) and their applications to the representation theory of the linear groups (see {\emph{e.g.}\cite{Macdo}).\\
The characters of the irreducible polynomial representations of the product group $G$ are product 
of Schur functions 
\[
S_{\lambda^{(1)},\dots,\lambda^{(k)}}=s_{\lambda^{(1)}}(X_1)\dots s_{\lambda^{(k)}}(X_k)
\]
where the $\lambda^{(i)}$ are partitions and each $X_i=\{x_{i1},\dots,x_{in_i}\}$ is a set of $n_i$ variables.
 The character of the one-dimensional representation 
\[
\det(g)^\ell=\det(g_1)^{\ell_1}\dots \det(g_k)^{\ell_k}.
\]
is the product of rectangular Schur functions $s_{\ell_1^{n_1}}(X_1)\dots s_{\ell_k^{n_k}}(X_k)$, whilst the character of $G$ is $s_d(X_1\dots X_k)$. Hence, the dimension of the space of invariants of degree $d$ and weight $\ell$, 
which is also the multiplicity of the one dimensional character $\det^\ell$ in $S^d(\mathcal H)$, is given by the scalar product 
\[
\dim {\mathrm Inv}(d,\ell)=\langle s_d(X_1\dots X_k),s_{\ell_1^{n_1}}(X_1)\dots s_{\ell_k^{n_k}}(X_k)\rangle_G
\]
of characters of $G$. To evaluate this scalar product, we can replace the $X_i$ by an infinite set of independent variables, and compute in the tensor product $Sym^{\otimes k}$ of $k$ copies of the algebra of symmetric functions $Sym$. The results will be the same, since in both cases the orthonormal basis is given by tensor products of Schur functions, $S_\lambda$ being identified with $s_{\lambda^{(1)}}\otimes \dots\otimes s_{\lambda^{(k)}}$.
Under this identification, the operation $\delta(f)=f(XY)$ corresponds to a 
comultiplication in $Sym$, which is known to be the adjoint of the internal product $\star$ of symmetric functions. 
Note the value of $\ell_i$ depends on whose of $d$ and $n_i$ ($\ell_in_i=d_i$).
Hence,
\begin{equation}\label{invdim}
\dim {\mathrm Inv}(d)=\langle s_d,s_{\ell_1^{n_1}}\star \dots \star s_{\ell_k^{n_k}}\rangle_{Sym}
\end{equation}
A similar reasoning gives
\begin{equation}\label{covdim}
\dim {\mathrm Cov}(d_0,\dots,d_k)=\langle s_{d_0},(s_{\ell_1^{n_1}}s_{d_1})\star \dots \star (s_{\ell_k^{n_k}}s_{d_k})\rangle_{Sym},
\end{equation}
again the value of each $\ell_i$ is obtained by $d_i+\ell_in_i=d_0$.\\

\noindent In order to compute the Hilbert series we need to introduce the Cauchy function $\Pi_t$ which is a very powerful tool for the manipulation of symmetric functions (see \emph{e.g.} \cite{Lasc}):
\[
\Pi_t(\mathbb X)=\prod_{x\in\mathbb X}\frac1{1-xt}=\exp\left\{\sum_{n\geq 1}t^n{p_n\over n}\right\}=\sum_{n\geq 0}s_nt^n
\]
where $p_n=\sum_{x\in\mathbb X}x^n$ denotes a power sum symmetric function.\\
Consider also the operator $$\check\partial_{t}:=\exp\left\{-\sum_{n\geq 1}(-t)^n{\partial\over \partial p_n}\right\}$$ and the vertex operator \cite{vertex}:
\[
\Gamma_t:=\Pi_t\check\partial_{-\frac1t},
\]
 The operator $\check\partial_{t}$ acts by shifting the power sum
\begin{equation}\label{shift}
\check\partial_{t}f(p_1,p_2,\dots)=f(p_1+t,p_2-t^2,\dots,).
\end{equation}
The definition of Schur function can be naturally extended to the compositions with negative parts \cite{Lasc}
\begin{equation}\label{schur}
s_v:=\left|\begin{array}{cccc}
s_{v_1}&s_{v_1+1}&\dots&s_{v_1+n-1}\\
s_{v_2}&s_{v_2+1}&\dots&s_{v_2+n-1}\\
\vdots&\dots&&\vdots\\
s_{v_n}&s_{v_n+1}&\dots&s_{v_n+n-1}\
 \end{array}\right|,
\end{equation}
with $s_n=0$ is $n<0$. Using equalities (\ref{shift}) and (\ref{schur}) we find 
\[
\Gamma_zs_v=\sum_{n\in\ZZ}z^ns_{nv}.
\]
And we use it iteratively to show 
\begin{equation}\label{vertex}
\sum_{\lambda_1,\dots,\lambda_{n-1}\in\ZZ\atop \lambda_n\in\NN} y_1^{\lambda_1}\dots y_n^{\lambda_n}s_{\lambda}=\prod_{i}y_i^{i-n}\prod_{i<j}(y_i-y_j)\Pi_1(YX),
\end{equation}
where $YX=\{y_ix:i=1\dots n,x\in X\}$.\\

On the other hand the dimension formula (\ref{covdim}) allows us to write the Hilbert series as a scalar product involving Cauchy functions. Let us denote by  $[\,]$ the plethysm operation (see \emph{e.g.} \cite{Macdo}). Remarking that for an alphabet of size $k$ we have: \[\sum_{l,d}s_{l^k}s_d=\sum_ns_n[s_1+s_{1^k}],\]
we obtain
\begin{equation}\label{series}
H_{Cov}(t;u)=\sum_{d_0,d_1,\dots,d_n}\dim {\mathrm Cov}(d_0,\dots,d_k)t^{d_0}u_1^{d_1}\dots u_n^{d_n}=
\langle \Pi_1[\alpha_{i_1}(t,u_1)],\Pi_1[\alpha_{i_2}(1,u_2)]\dots\Pi_1[\alpha_{i_k}(1,u_n)]\rangle
\end{equation}
where $\alpha_{k}(t,u)=tus_1+t^ks_{1^k}$ for $k$-ary alphabet.  Note we have
\[
\Pi_1[us_1+s_{1^k}]=\sum_{\lambda=(\lambda_1,\dots,\lambda_k)\atop \lambda_1\geq \lambda_2=\dots=\lambda_k}u^{\lambda_1-\lambda_2}s_\lambda;
\]
this is a consequence of $$s_n[s_{1^k}]=\sum_{\lambda=(\lambda_1,\dots,\lambda_p)\atop\lambda_1+\dots+\lambda_k=n}s_{\lambda_1^k\dots\lambda_p^k}$$ and of the Pieri formula.
Hence, for an alphabet of size $k$ we have
\[
\Pi_1[us_1+s_{1^k}]=\sum_{\lambda=(\lambda_1,\dots,\lambda_k)\atop \lambda_1\geq \lambda_2=\dots=\lambda_k}u^{\lambda_1-\lambda_2}s_\lambda.
\]

These series can be obtained using a combination of vertex operators and Omega operators of Macmahon $\Omega^u_{\geq}$ \cite{MacMa} which send the monomials with a negative power of $u$ to $0$. Indeed, from (\ref{vertex}) we obtained for an alphabet of size $2$ (binary case)
\[
\sum_{\ell(\lambda)\leq 2}u^{\lambda_1-\lambda_2}s_\lambda=\Omega^u_{\geq}\left(1-{1\over u^2}\right)\Pi_t\left[\left(u+\frac1u\right)\mathbb X\right] 
\]
and for an alphabet of size $3$ (ternary case), by setting $y_1=u$, $y_2=\frac 1{vu}$ and $y_3=v$ in (\ref{vertex})
\[\begin{array}{rcl}\displaystyle
\sum_{\lambda=(\lambda_1,\lambda_2,\lambda_2)\in\NN^3\atop
\lambda_1\geq\lambda_2}u^{\lambda_1-\lambda_2}s_\lambda&=&
\displaystyle{\rm CT}_v\Omega^u_{\geq}\sum_{\lambda=(\lambda_1,\lambda_2,\lambda_3)\atop \lambda_1,\,\lambda_2\in\ZZ,\,\lambda_3\in\NN}u^{\lambda_1-\lambda_2}v^{\lambda_2-\lambda_3}s_\lambda
\\&=&\displaystyle{\rm CT}_v\Omega^u_{\geq}\left(1-{1\over vu^2}\right)
\left(1-{v\over u}\right)\left(1-{1\over v^2u}\right)\Pi_t\left[\left(u+\frac 1{vu}+v\right)\mathbb X\right],
\end{array}
\]
where $\mathrm{CT}_v$ means the constant term.\\
Combining this with equality (\ref{series}) this gives the Hilbert series as an Omega
\begin{equation}\label{HilbForm}
H_{Cov}(t;u)={\rm CT}_{v_1}\Omega^{u_1}_{\geq}\dots{\rm CT}_{v_n}\Omega^{u_n}_{\geq}B_{i_1}(u_1,v_1)\dots B_{i_n}(u_n,v_n)\Pi_t\left[A_{i_1}(u_1,v_1)\cdots A_{i_n}(u_n,v_n)\right]
\end{equation}
where  $A_2(u,v)=u+\frac1{u}$, $B_2(u,v)=1-{1\over u^2}$ for binary variables and $$A_2(u,v)=u+v+\frac 1{vu},\, B_3(u,v)=\left(1-{1\over vu^2}\right)\left(1-{1\over v^2u}\right)
\left(1-{v\over u}\right)$$ for ternary variables.
\section{Interpretations in terms of annalagmatic geometry\label{circles}}

 Let us recall first the principle of tetracyclic coordinates \cite{Cool}.\\
One starts with the equation of a
circle in the real Euclidean plane with coordinates $(x,y)$.
Introducing homogenous coordinates $(X:Y:Z)$, on the complexified
plane, the equation of our circle $C$ can be written in the form
\begin{equation}
x_0\cdot  i(X^2+Y^2+Z^2)+x_1\cdot(X^2+Y^2-Z^2)+ x_2\cdot 2XZ + x_3
\cdot 2YZ=0\,,
\end{equation}
where $(x_0:x_1:x_2:x_3)$ are called the (homogeneous) coordinates
of $C$. The quantities $y_0= i(X^2+Y^2+Z^2)$, $y_1= X^2+Y^2-Z^2$,
$y_2=2XZ$, $y_3=2YZ$, are called the {\it special tetracyclic
coordinates} of the point $(X:Y:Z)$ of $\PP^2$.
They satisfy
\begin{equation}
(yy)=0\,,
\end{equation}
where
\begin{equation}
(xy)=\sum_{i=0}^3 x_iy_i
\end{equation}
is the fundamental quadratic form in the geometry of circles. Any
other nondegenerate quadratic form can be taken instead of $(xy)$.
We shall set for a circle
\begin{equation}
x_0=a+d\,,\ x_1=i(a-d)\,,\ x_2=b-c\,\ x_3=i(b+c)\,,
\end{equation}
so that
\begin{equation}
(xx)=2(ad-bc) = 2\det M\,\quad (xx')=ad'+a'd-bc'-b'c
\end{equation}
where $M$ is the matrix
\begin{equation}
M=\begin{pmatrix} a&b\\ c&d \end{pmatrix}.
\end{equation}
The tetracyclic coordinates of a point is now given by a rank 1
matrix which can be written in the form
\begin{equation}
\begin{pmatrix} a&b\\ c&d \end{pmatrix}
=\begin{pmatrix} y_2 z_2 & -y_2z_1 \\ -y_1z_2 &
y_1z_1\end{pmatrix}\,.
\end{equation}
In this picture, finite points of the plane are represented by
points on the quadric $\Sigma$: $\det M=0$, and the coordinates of
a point are given in terms of the parameters $(y_1,y_2)$ and
$(z_1,z_2)$ of the two generatrices of the quadric intersecting in
it. The finite points are those for which $ix_0+x_1=-2Z^2\not=0$.
In matrix coordinates, this reads $a\not=0$. Hence, the elements
at infinity of this geometry consist of the two intersecting
generatrices $b=0$ and $c=0$ of $\Sigma$. We are working with a
compactification of the complexified Euclidean plane $\CC^2$ which
is not isomorphic to $\PP^2$ but to $\PP^1\times\PP^1$. The
generatrices of $\Sigma$ correspond to two families of lines of
the affine plane, called {\it minimal lines}.

\noindent Turning back to the multilinear form $2\times2\times2$, we can now interpret the
equation $\sum_{i,j=0}^1a_{ij}x_iy_j=0$ as a that of a circle, whose coordinates form the
matrix $(a_{ij})$. Hence, the equation
$A=0$ can be understood a describing a pencil of
circles. To do this, we have to single out one variable, say
$z$, viewed as a projective parameter, the other two ones being
minimal line coordinates on a tetracyclic plane. In this setting,
the orbit classification and the normal forms are almost immediate.
Let $\mathcal C_z$ be the point in $\PP^3$ representing the circle of the
pencil with parameter $z$. Then, the set
$\ell=\{\mathcal C_z|z\in\CC^2\}$ can be either a proper line (the generic
case) or be degenerated into a single point. The rest of the
discussion will depend on the relative position of this line or
point with respect to the non-singular quadric $\Sigma$ of null
circles. Let us first consider the case where $\ell$ is a proper
line. Generically, it will intersect  $\Sigma$ into two distinct
points $\mathcal C_1$, $\mathcal C_2$ (the base points of the pencil). These points
can be mapped by a circular transformation to the origin of the
affine plane, and to the intersection of the two isotropic lines
at infinity, which have respectively as matrix coordinates
\begin{equation}
\begin{pmatrix} 1&0\\0&0\end{pmatrix}\quad {\rm and}\quad
\begin{pmatrix}0&0\\0&1\end{pmatrix}
\end{equation}
whence the normal form associated to $\mathcal O_{VI}$ in this case. The test for this case
is $\Delta\not=0$, since $B_z=0$ gives the parameters for the two
null circles of the pencil. We note on the normal forms that
$C=0$ is a circle of the same pencil. It is the
unique such circle which is orthogonal to $\mathcal C_z$. Let us remark
that since we are working over $\C$, we do not distinguish between
intersecting pencils and pencils with limit points, but it would
be easy to refine the discussion in order to obtain the
classification over $\R$, which can be found in Sokolov's book
\cite{Soko}.

\noindent The next case is when $\ell$ is tangent to $\Sigma$ in exactly one
point. Here, $\Delta=0$, and there is exactly one null circle
$\mathcal C_z$. By a circular transformation, we can arrange that this
null circle becomes the origin, and that the radical axis becomes
a  coordinate axis, say the $y$ axis whose equation in matrix
tetracyclic coordinates is $b-c=0$, $d=0$, whence the normal form associated to
$\mathcal O_{V}$ for this case. All the circles of the pencil have its null
circle as a common  point. Its equation is given by $C=0$.

\noindent By a further degeneracy, $\ell$ can become a generatrix of $\Sigma$.
The two systems constitute then two orbits, with respective normal
forms associated to $\mathcal O_{II}$ and $\mathcal O_{III}$.

\noindent Finally, $\ell$ can be a single point $m$. If it is not on $\Sigma$,
we can transform it into any proper circle, e.g., the one with
matrix coordinates $$\begin{pmatrix} 0&1\\1&0\end{pmatrix}\,, $$
which yields the normal form associated to $\mathcal O_{IV}$. If $m$ is on $\Sigma$, it
is a null circle, and the normal form can obviously be taken as
whose associated to $\mathcal O_{I}$.

\noindent For the case $2\times2\times3$: the equation $A=0$ represents then a linear complex
(or {\it net}) of circles $\mathcal C_z$, i.e. a plane of the $\PP^3$ of
circles, which may degenerate into a line or a point. Let, as
above, $\Pi_x$ be the linear variety formed by the representative
points of all these circles. The ground form, identified with the matrix
 $M_z=\left(a_{ij0}z_0+a_{ij1}z_1+a_{ij2}z_0\right)_{i,j=0,1}=\sum_iM_iz_i$, is then the
matrix of circle coordinates of $\mathcal C_z$, and
$B=\det(M_z)=0$ when $\mathcal C_z$ is a null circle.

\noindent In the generic case, $M_1$, $M_2$, $M_3$ are linearly independent,
and we have a proper net. This case is recognized from the
covariant $C$, which does not vanish identically. The net
is then formed by the collection of circles orthogonal to a fixed
circle, whose equation is $C=0$. By a circular
transformation, this circle  can be mapped to a coordinate axis,
say $0z$, whence a simple normal form (the net of circles centered
on $0z$). This is not anymore possible if $\det C=0$. The net is
the formed by all circle having a common point, which me may take
as the origin of coordinates.

\noindent If $C$ vanishes identically, the net reduces to a pencil or to a
point, and the normal forms can be inferred from the previous
discussion of trilinear forms.

\noindent To conclude, let us remark that $D_xD_y=0$ gives the minimal lines
through the common points of $A=0$ and $C=0$. Also, the
hyperdetrminant $\Delta$ is proportional to the discriminant of
$B$, whose determinantal expression is recognized as the
condition that the three circles with coordinates $M_i$ have a
common point.

\noindent{\bf Acknowledgments} This paper is partially supported by the ANR project PhysComb,
ANR-08-BLAN-0243-04.

\end{document}